\documentclass[ALICE,manyauthors]{cernphprep}
\usepackage[comma,square,numbers,sort&compress]{natbib}
\usepackage{hyperref}
\usepackage{xcolor}
\usepackage{lineno}
\usepackage{xspace}
\usepackage{multirow}
\usepackage{floatrow}
\usepackage[T1]{fontenc}
\usepackage{orcidlink}

\begin{document}
%

\newcommand{\pp}           {pp\xspace}
\newcommand{\ppbar}        {\mbox{$\mathrm {p\overline{p}}$}\xspace}
\newcommand{\XeXe}         {\mbox{Xe--Xe}\xspace}
\newcommand{\PbPb}         {\mbox{Pb--Pb}\xspace}
\newcommand{\pA}           {\mbox{pA}\xspace}
\newcommand{\pPb}          {\mbox{p--Pb}\xspace}
\newcommand{\AuAu}         {\mbox{Au--Au}\xspace}
\newcommand{\dAu}          {\mbox{d--Au}\xspace}

\newcommand{\s}            {\ensuremath{\sqrt{s}}\xspace}
\newcommand{\snn}          {\ensuremath{\sqrt{s_{\mathrm{NN}}}}\xspace}
\newcommand{\pt}           {\ensuremath{p_{\rm T}}\xspace}
\newcommand{\meanpt}       {$\langle p_{\mathrm{T}}\rangle$\xspace}
\newcommand{\ycms}         {\ensuremath{y_{\rm CMS}}\xspace}
\newcommand{\ylab}         {\ensuremath{y_{\rm lab}}\xspace}
\newcommand{\etarange}[1]  {\mbox{$\left | \eta \right |~<~#1$}}
\newcommand{\yrange}[1]    {\mbox{$\left | y \right |~<~#1$}}
\newcommand{\dndy}         {\ensuremath{\mathrm{d}N_\mathrm{ch}/\mathrm{d}y}\xspace}
\newcommand{\dndeta}       {\ensuremath{\mathrm{d}N_\mathrm{ch}/\mathrm{d}\eta}\xspace}
\newcommand{\avdndeta}     {\ensuremath{\langle\dndeta\rangle}\xspace}
\newcommand{\dNdy}         {\ensuremath{\mathrm{d}N_\mathrm{ch}/\mathrm{d}y}\xspace}
\newcommand{\Npart}        {\ensuremath{N_\mathrm{part}}\xspace}
\newcommand{\Ncoll}        {\ensuremath{N_\mathrm{coll}}\xspace}
\newcommand{\dEdx}         {\ensuremath{\textrm{d}E/\textrm{d}x}\xspace}
\newcommand{\RpPb}         {\ensuremath{R_{\rm pPb}}\xspace}

\newcommand{\nineH}        {$\sqrt{s}~=~0.9$~Te\kern-.1emV\xspace}
\newcommand{\seven}        {$\sqrt{s}~=~7$~Te\kern-.1emV\xspace}
\newcommand{\twoH}         {$\sqrt{s}~=~0.2$~Te\kern-.1emV\xspace}
\newcommand{\twosevensix}  {$\sqrt{s}~=~2.76$~Te\kern-.1emV\xspace}
\newcommand{\five}         {$\sqrt{s}~=~5.02$~Te\kern-.1emV\xspace}
\newcommand{\twosevensixnn}{$\sqrt{s_{\mathrm{NN}}}~=~2.76$~Te\kern-.1emV\xspace}
\newcommand{\fivenn}       {$\sqrt{s_{\mathrm{NN}}}~=~5.02$~Te\kern-.1emV\xspace}
\newcommand{\LT}           {L{\'e}vy-Tsallis\xspace}
\newcommand{\GeVc}         {Ge\kern-.1emV$/c$\xspace}
\newcommand{\MeVc}         {Me\kern-.1emV/$c$\xspace}
\newcommand{\TeV}          {Te\kern-.1emV\xspace}
\newcommand{\GeV}          {Ge\kern-.1emV\xspace}
\newcommand{\MeV}          {Me\kern-.1emV\xspace}
\newcommand{\GeVmass}      {Ge\kern-.2emV/$c^2$\xspace}
\newcommand{\MeVmass}      {Me\kern-.2emV/$c^2$\xspace}
\newcommand{\lumi}         {\ensuremath{\mathcal{L}}\xspace}

\newcommand{\ITS}          {\rm{ITS}\xspace}
\newcommand{\TOF}          {\rm{TOF}\xspace}
\newcommand{\ZDC}          {\rm{ZDC}\xspace}
\newcommand{\ZDCs}         {\rm{ZDCs}\xspace}
\newcommand{\ZNA}          {\rm{ZNA}\xspace}
\newcommand{\ZNC}          {\rm{ZNC}\xspace}
\newcommand{\SPD}          {\rm{SPD}\xspace}
\newcommand{\SDD}          {\rm{SDD}\xspace}
\newcommand{\SSD}          {\rm{SSD}\xspace}
\newcommand{\TPC}          {\rm{TPC}\xspace}
\newcommand{\TRD}          {\rm{TRD}\xspace}
\newcommand{\VZERO}        {\rm{V0}\xspace}
\newcommand{\VZEROA}       {\rm{V0A}\xspace}
\newcommand{\VZEROC}       {\rm{V0C}\xspace}
\newcommand{\Vdecay} 	   {\ensuremath{V^{0}}\xspace}

\newcommand{\ee}           {\ensuremath{\rm e^{+} \rm e^{-}}} 
\newcommand{\pip}          {\ensuremath{\pi^{+}}\xspace}
\newcommand{\pim}          {\ensuremath{\pi^{-}}\xspace}
\newcommand{\kap}          {\ensuremath{\rm{K}^{+}}\xspace}
\newcommand{\kam}          {\ensuremath{\rm{K}^{-}}\xspace}
\newcommand{\pbar}         {\ensuremath{\rm\overline{p}}\xspace}
\newcommand{\kzero}        {\ensuremath{{\rm K}^{0}_{\rm{S}}}\xspace}
\newcommand{\lmb}          {\ensuremath{\Lambda}\xspace}
\newcommand{\almb}         {\ensuremath{\overline{\Lambda}}\xspace}
\newcommand{\Om}           {\ensuremath{\Omega^-}\xspace}
\newcommand{\Mo}           {\ensuremath{\overline{\Omega}^+}\xspace}
\newcommand{\X}            {\ensuremath{\Xi^-}\xspace}
\newcommand{\Ix}           {\ensuremath{\overline{\Xi}^+}\xspace}
\newcommand{\Xis}          {\ensuremath{\Xi^{\pm}}\xspace}
\newcommand{\Oms}          {\ensuremath{\Omega^{\pm}}\xspace}
\newcommand{\degree}       {\ensuremath{^{\rm o}}\xspace}
\floatsetup[table]{capposition=top}

\newcommand{\py}{PYTHIA~8\xspace}
\newcommand{\pya}{PYTHIA~8.244\xspace}
\newcommand{\ep}{EPOS LHC\xspace}
\newcommand{\ptt}{\ensuremath{p_{\mathrm{T}}^{\rm trig}}\xspace}
\newcommand{\pta}{\ensuremath{p_{\mathrm{T}}^{\rm assoc}}\xspace}
\newcommand{\nch}{\ensuremath{N_{\mathrm{ch}}^{\mathrm{T}}}\xspace}
\newcommand{\nacc}{\ensuremath{N_{\mathrm{acc}}^{\mathrm{T}}}\xspace}
\newcommand{\nraw}{\ensuremath{N_{\mathrm{raw}}^{\mathrm{T}}}\xspace}
\newcommand{\rt}{\ensuremath{R_{\mathrm{T}}}\xspace}
\newcommand{\mpt}{\ensuremath{\langle p_{\mathrm{T}} \rangle}\xspace}
\newcommand{\va}{\ensuremath{\varphi^{\mathrm{assoc}}}\xspace}
\newcommand{\vt}{\ensuremath{\varphi^{\mathrm{trig}}}\xspace}
\newcommand{\red}[1]{\textcolor{red}{#1}}
\newcommand{\blue}[1]{\textcolor{blue}{#1}}
\newcommand{\green}[1]{\textcolor{green}{#1}}

\begin{titlepage}
\PHyear{2023}       
\PHnumber{212}      
\PHdate{20 September}  

\title{Charged-particle production as a function of the relative transverse activity classifier in \pp, \pPb, and \PbPb collisions at the LHC}
\ShortTitle{Charged-particle production as a function of \rt at the LHC}   

\Collaboration{ALICE Collaboration\thanks{See Appendix~\ref{app:collab} for the list of collaboration members}}
\ShortAuthor{ALICE Collaboration} 

\begin{abstract}

Measurements of charged-particle production in \pp, \pPb, and \PbPb collisions in the toward, away, and transverse regions with the ALICE detector are discussed.  These regions are defined event-by-event relative to the azimuthal direction of the charged trigger particle, which is the reconstructed particle with the largest transverse momentum (\ptt) in the range $8<\ptt<15$\,GeV$/c$. The toward and away regions contain the primary and recoil jets, respectively; both regions are accompanied by the underlying event (UE). In contrast, the transverse region perpendicular to the direction of the trigger particle is dominated by the so-called UE dynamics, and includes also contributions from initial- and final-state radiation. The relative transverse activity classifier, $\rt=\nch/\langle\nch\rangle$, is used to group events according to their UE activity, where \nch is the charged-particle multiplicity per event in the transverse region and $\langle\nch\rangle$ is the mean value over the whole analysed sample.  The energy dependence of the \rt distributions in \pp collisions at $\s=2.76$, 5.02, 7, and 13\,TeV is reported, exploring the Koba-Nielsen-Olesen (KNO) scaling properties of the multiplicity distributions. The first measurements of charged-particle \pt spectra as a function of \rt in the three azimuthal regions in \pp, \pPb, and \PbPb collisions at $\sqrt{s_{\rm NN}}=5.02$\,TeV are also reported. Data are compared with predictions obtained from the event generators \py and \ep. This set of measurements is expected to contribute to the understanding of  the origin of collective-like effects in small collision systems (\pp and \pPb).  

\end{abstract}
\end{titlepage}

\setcounter{page}{2} 


\section{Introduction} 
\begin{figure}
        \centering
       \includegraphics[width=6.5cm, height=8cm]{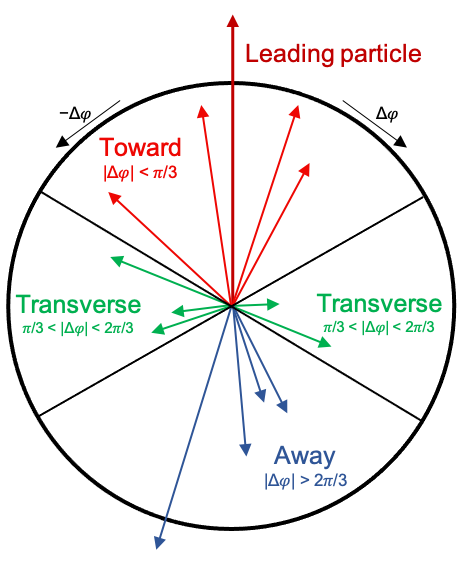}
        \caption{Schematic representation of the toward, transverse, and away regions in the azimuthal plane with respect to the leading particle, i.e.\,the particle with the highest \pt in the event. Figure taken from Ref.~\cite{ALICE:2022qxg}.}
        \label{TR}
    \end{figure}

    In proton--proton (\pp) collisions in which a partonic scattering with large momentum transfer occurs, along with the particles originating from the hadronisation of the parton showers initiated by the hard-scattered partons (jets), also low-\pt particles from proton break-up (``proton remnants'') and multi-parton interactions (MPI), are produced~\cite{ALICE:2022fnb}. This collection of low-\pt particles is termed as underlying event (UE). Understanding and accurate modelling such components is important to ensure a proper description of particle production in nuclear collisions. To this end, the kinematic region containing the fragmentation products of the main partonic scattering needs to be separated from the remaining UE part~\cite{ALICE:2022qxg}. Experimentally, it is impossible to uniquely separate the UE from the event-by-event hard scattering process. However, the UA1 experiment in proton--antiproton collisions at CERN perfomed the first study of this kind by measuring the transverse energy density outside the leading jet, which is also known as jet pedestal region~\cite{UA1:1983hhd,UA1:1988sgw,UA1:1989bou}. Another approach based on the definition of three distinct topological regions was introduced by the CDF Collaboration~\cite{CDF:2001onq}. The three topological regions are defined from the angular difference between the trigger and associated particles, $|\Delta \varphi| = \va - \vt$, where \vt and \va refer to the value of the azimuthal angle for the trigger particle and for associated particles in the event, respectively~\cite{ALICE:2019mmy}. The trigger particle is the one with the largest transverse momentum (\ptt) in the event, and the rest are termed as associated particles. The criteria for the definition of the different topological regions are depicted in Fig.~\ref{TR}.  The toward region ($|\Delta \varphi|<\pi/3$) contains the main jet while the away region ($|\Delta \varphi|>2\pi/3$) may involve the fragments of the recoil jet. In general, these two regions are less sensitive to the UE. In contrast, the transverse region ($\pi/3$ $<|\Delta \varphi|<2\pi/3$) is less affected by contributions from the hard scattering. However, this region is expected to contain particles from initial- and final-state radiation (ISR and FSR)~\cite{CDF:2001onq}. Moreover, the multiplicity density in the transverse region in events with jets is known to be about twice the multiplicity density in minimum bias collisions due to a centrality bias induced by the presence of a hard scattering, which restricts the impact parameter fluctuations~\cite{ALICE:2019mmy}.

    Recently, data at high multiplicity from \pp collisions at RHIC~\cite{Adam_2020} and from \pp and \pPb collisions at the LHC~\cite{ALICE:2016fzo,ALICE:2013wgn} have shown striking similarities with heavy-ion data, such as collectivity and strangeness enhancement, which could be explained by the formation of a strongly-coupled quark--gluon plasma (sQGP)~\cite{Busza:2018rrf,Nagle:2018nvi}. In order to understand the origin of these collective-like effects in pp collisions, it is important to use observables sensitive to them and compare the data with models such as \py~\cite{Sjostrand:2014zea} and \ep~\cite{Pierog:2013ria}. In \py, mechanisms like colour reconnection~\cite{OrtizVelasquez:2013ofg}, rope hadronisation~\cite{Bierlich:2014xba}, and string shoving~\cite{Bierlich:2020naj} can produce effects resembling those from a collective behaviour, while the core--corona approach implemented in \ep, including a collective expansion for the core, can describe some aspects of the data in \pp and \pPb collisions. Despite the similarities between the results from small collision system (\pp and \pPb collisions) and heavy-ion collisions, one of the major difficulties in getting a clear picture is to understand the selection biases and autocorrelation effects in  different collision systems. By selecting a high event activity (high particle multiplicity) in small collision system, the event sample is naturally biased towards hard processes~\cite{ALICE:2019mmy}. To overcome such selection biases and get an insight on how these biases affect the charged particle \pt spectra, the transverse region can be used to build a new event classifier, \rt, which is expected to have low sensitivity to the  hard processes~\cite{Martin:2016igp}.  The relative transverse activity classifier, \rt, is the ratio of the primary charged-particle multiplicity in the transverse region (\nch) obtained event-by-event to the average value ($\langle \nch \rangle$) ~\cite{Martin:2016igp,ALICE:2019mmy}. It is defined as
    
    \begin{equation}\rt=\frac{N^{\mathrm{T}}_{\mathrm{ch}}}{\langle N^{\mathrm{T}}_{\mathrm{ch}} \rangle}.
        \label{Rt}
    \end{equation}

    Although the purpose of \rt is to quantify the UE activity of the events, it can still be influenced by jet fragments, especially for events with high \rt values in experiments with limited acceptance~\cite{Bencedi:2020qpi,Bencedi:2021tst}.

    For small systems, \rt can help to reveal whether the properties of events with lower UE contribution (``low-UE") are compatible with equivalent measurements in \ee collisions (jet universality), and whether the scaling behaviour of events with higher UE contribution (``high-UE") exhibits properties of non-trivial soft-QCD dynamics, such as colour reconnection or other phenomena~\cite{Martin:2016igp,Ortiz:2017jaz,Ortiz:2018vgc}. The event classifier \rt has been recently used in the analysis of identified charged-particle production in \pp collisions at $\sqrt{s}=13$\,TeV at the LHC~\cite{ALICE:2023yuk}. In this paper, \rt is used for the first time for \pPb and \PbPb collisions as well as \pp collisions at lower centre-of-mass energies. The application of \rt to \pPb and \PbPb data would provide a new set of multidimensional measurements, which can serve  to improve the theoretical modelling of the complex interplay of hard and soft QCD processes across system sizes. This is particularly relevant for the new developments in PYTHIA~8~\cite{Bierlich:2018xfw}.
    
    The energy dependence of the \rt distributions in \pp collisions at centre-of-mass energies $\sqrt{s}=2.76$, 5.02, 7, and 13\,TeV is studied in this paper, exploring the KNO~\cite{Koba:1972ng} scaling of the multiplicity distributions in the transverse region~\cite{Ortiz:2017jaz}. The KNO scaling is the hypothesis that at high \s the probability distributions $P(n)$ of producing $n$ particles in a certain collision process should exhibit the scaling relation
    \begin{equation} \label{eq:4}
        P(n) = \frac{1}{\langle n\rangle}\Psi\left(\frac{n}{\langle n\rangle}\right),
    \end{equation}
    with $\langle n\rangle$ being the average multiplicity. This means that after scaling with $\langle n\rangle$, measured $P(n)$ at different energies collapse onto a universal function $\Psi$~\cite{Hegyi:2000sp}. The KNO scaling is expected in models which assume that a single \pp collision is merely a superposition of a given number of elementary partonic collisions emitting particles independently~\cite{DiasdeDeus:1997ui}. Therefore, multi-parton interactions are expected to produce such an effect~\cite{Ortiz:2017jaz}. 
    
    Also, a study on \pt-spectra of primary charged particles as a function of \rt in \pp, \pPb, and \PbPb collisions at a centre-of-mass energy per nucleon pair of $\sqrt{s_{\rm NN}} = 5.02$\,TeV on the \rt distributions is presented.
    The \pt spectra are studied in the toward, away, and transverse regions.  The results from \pp collisions are compared with predictions from \py with Monash tune~\cite{Skands:2014pea} (from now on referred to as \py) and \ep. For \pPb and \PbPb collisions, data are compared with \ep and \py/Angantyr~\cite{Bierlich:2018xfw}.

    In \pp collisions, the implementation of hard processes in \py starts with the choice of the parton distribution functions (PDFs) of the protons. The next step is to simulate the hard scattering process, where partons from the colliding protons interact at high energies. This process is described using perturbative quantum chromodynamics (pQCD) calculations~\cite{Sjostrand:2014zea}.  The event generation begins with a fixed order matrix element calculation either in leading-order (LO), next-to-leading order (NLO) or beyond (NNLO)~\cite{Skands}. During the hard scattering process, partons may emit ISR and FSR which are simulated using parton shower algorithms~\cite{Sjostrand:2014zea}. \py with Monash tune has a default parameterisation of the model based on MPI and colour reconnection. After the hard scattering and parton showering, coloured strings are formed between the final-state partons~\cite{Ortiz_2017}. The hadronisation mechanism in \py is based on the Lund string fragmentation model~\cite{Sjostrand:2014zea}, followed by particle decays. The hadronisation process leads to the production of jets and the UE. The Angantyr model in \py is an extrapolation of the \pp dynamics to collisions with nuclei with a minimal set of tunable parameters that tries to describe the general features of the final state in p--A and A--A collisions, such as multiplicity and transverse momentum distributions, without including a hydrodynamic evolution. It is based on the Fritiof model and the number of participating nucleons is calculated using the Glauber formalism~\cite{Bierlich:2018xfw, ALICE:2022fnb}. On the other hand, the event generation process in \ep arises from the parton-based Gribov--Regge theory~\cite{2019drk}, pQCD, and the Lund string model. An elementary scattering corresponds to a scattering of primary partons, which contains a hard scattering (pQCD), accompanied by ISR and FSR. String hadronisation relies on the local density of string segments per volume unit with respect to a critical density parameter. Each string is classified according to a core--corona approach where a low-density corona region accompanies a high-density core region~\cite{2019drk}. In the core, the string energy density is sufficient to invoke a QGP description that is subject to hydrodynamic evolution~\cite{ALICE:2022wpn}.
    
    This article is structured as follows: Sec.~\ref{sec:details} is devoted to the discussion of the main aspects of the analysis, such as event and track selection, corrections, systematic uncertainties as well as the unfolding method for the \rt distributions and \pt spectra. The results are discussed in Sec.~\ref{sec:results}, and finally Sec.~\ref{sec:conclusion} summarises the main results.

\section{Analysis procedure}
\label{sec:details}
\subsection{Event and track selection}
    The present study is performed using the LHC Run~1 and Run~2 data collected with the ALICE detector. The Inner Tracking System (\ITS~\cite{ALICE:1999cls}), Time Projection Chamber (\TPC~\cite{ALICE:2000jwd}), and \VZERO~\cite{ALICE:2004ftm} are the main detectors used in the current analysis. The \ITS is composed of six cylindrical layers of high-resolution silicon tracking detectors. The two innermost layers, closest to the interaction point (IP), consist of hybrid Silicon Pixel Detectors (\SPD) at radial distances of 3.9 and 7.6\,cm from the beam axis with a pseudorapidity coverage of $|\eta|<2$ and $|\eta|<1.4$, respectively. The \TPC is a cylindrical drift detector that covers a radial distance of 85--247\,cm from the beam axis and its longitudinal dimension extends from about $-250$\,cm to $+250$\,cm around the nominal IP. The \VZERO detector consists of two arrays of scintillating counters (\VZEROA and \VZEROC) placed on each side of the IP covering the full azimuthal acceptance and the pseudorapidity ranges of $2.8<\eta<5.1$ and $-3.7<\eta<-1.7$, respectively~\cite{ALICE:2008aamodt}. The \VZERO detector and the \SPD are used for triggering and background rejection. The data were collected using a minimum-bias (MB) trigger, which required a signal in both \VZEROA and \VZEROC detectors.  
    The offline event selection is optimised to reject beam-induced background in all collision systems by utilising the timing signals in the two \VZERO detectors. In \PbPb collisions, in order to suppress the beam induced background and the electromagnetic interactions, the \VZERO-timing selection is complemented by correlating the timing signals of the neutron zero degree calorimeters (\ZDC~\cite{ALICE:1999edx}), which are positioned on both sides of the IP at 112.5\,m distance along the beam axis~\cite{ALICE:2014sbx}. Runs with a low number of interactions per bunch crossing ($\mu$) were selected resulting in average $\mu$ values of 0.020, 0.005, and 0.001 for \pp, \pPb, and \PbPb collisions at $\sqrt{s_{\rm NN}}=5.02$\,TeV, respectively. Therefore, events with multiple collisions (pile-up) constitute a small fraction of the triggered events. They are identified and rejected based on the presence of multiple offline reconstruction primary vertices in the \SPD~\cite{ALICE:2008aamodt,ALICE:2018pal,ALICE:2019mmy}. To ensure that a hard scattering took place in a collision, events are required to have a trigger particle with transverse momentum above a given threshold. For the energy dependence of the \rt distributions in pp collisions, the trigger particle is chosen in the same \pt interval (5--40\,GeV$/c$) as used in previous publications for pp collisions at $\sqrt{s}=13$\,TeV~\cite{ALICE:2019mmy,ALICE:2023yuk}. However, for the system size dependence of \pt spectra as a function of \rt, \pp, \pPb, and \PbPb collisions with a trigger particle within $8<\ptt<15$\,GeV$/c$ are considered. This choice is motivated from previous ALICE results where the particle production in the toward and away regions are studied as a function of  the collision centrality using heavy-ion data~\cite{ALICE:2011gpa,ALICE:2022qxg}. For \pPb and \PbPb collisions, in this particular \ptt interval the jet-like correlations dominate over the collective effects, and therefore, the separation in three topological regions is appropriate~\cite{ALICE:2011gpa}. 
    
    This work is based on the analysis of ALICE data including \pPb and \PbPb collisions at $\snn = 5.02$\,TeV and \pp collisions at $\s=2.76$, 5.02, 7, and 13\,TeV. The measurements of the transverse momentum spectra focus on primary charged particles~\cite{ALICE:2017prim}, i.e. particles with a mean proper lifetime larger than 1\,cm$/c$, which are either produced directly in the interaction or from decays of particles with mean proper lifetime smaller than 1\,cm$/c$. Primary charged particles are measured in the pseudorapidity range of $|\eta|<0.8$ and with $\pt > 0.5$\,\GeVc. They are reconstructed using the ITS and TPC detectors, which provide measurements of the transverse momentum of the track and its azimuthal angle. 
    
    For the measurement of the \pt spectra as a function of \rt, the track selection criteria are similar to those used in the \rt studies for identified charged particles reported in Ref.~\cite{ALICE:2023yuk}. In particular, tracks are required to cross at least 70 TPC pad rows. They are also required to have at least two hits in the ITS, out of which at least one has to be from track segments in the SPD layers. The fit quality for the ITS and TPC track points must satisfy $\chi^{2}_{\rm ITS}/N_{\rm hits} < 36$ and $\chi^{2}_{\rm TPC}/N_{\rm clusters} < 4$, respectively, where $N_{\rm hits}$ and $N_{\rm clusters}$ are the number of hits in the ITS and the number of clusters in the TPC associated with the track, respectively. To limit the contamination from secondary particles, a selection on the distance of closest approach (DCA) to the reconstructed primary vertex in the direction parallel to the beam axis ($z$) of $|{\rm DCA}_{z}| < 2$\,cm is applied. Also, a \pt-dependent selection on the DCA in the transverse plane (${\rm DCA}_{xy}$) of the selected tracks to the primary vertex is applied ($|{\rm DCA}_{xy}|<0.0105$\,cm + $0.0350$\,cm$\times({\rm GeV}/c)^{-C}\times p_{\rm T}^{C}$ with $p_{\rm T}$ in ${\rm GeV}/c$ and $C=-1.1$)~\cite{ALICE:2022qxg}. Moreover, tracks associated with the decay products of weakly decaying kaons (``kinks'') are rejected. These track selection criteria yield a significantly non-uniform efficiency as a function of the azimuthal angle and the pseudorapidity. This is mostly due to the requirement of SPD hits. In order to obtain a high and uniform tracking efficiency together with good momentum resolution, they are complemented by tracks without an associated hit in the SPD for which the position of the reconstructed primary vertex is used in the fit of the tracks~\cite{ALICE:2012eyl,ALICE:2012nbx}. Both sets of tracks are used to select the trigger particle as well as to measure \rt and the \pt spectra.

    On the other hand, for the measurement of the \rt distributions in \pp collisions similar track selection criteria as described in Ref.~\cite{ALICE:2019mmy} were considered. However, in order to guarantee a uniform response in the azimuth, the DCA cut was loosen ($|{\rm DCA}_{xy}|<2.4$) and no restriction on the number of reconstructed points in ITS was considered. The obtained results are consistent with those using the combination of tracks selected with the two sets of criteria described above.

\subsection{Corrections}
     For the measurements of the \pt spectra of charged particles, the standard procedure of the ALICE Collaboration considers efficiency and secondary particle contamination to correct the raw yields~\cite{ALICE:2018vuu}. The efficiency correction is calculated from Monte Carlo (MC) simulations, including particle propagation through the detector making use of the GEANT 3 transport code~\cite{Brun:1082634}. The event generators used in the analyses are \py for \pp collisions, \ep~\cite{Pierog:2013ria} for \pPb collisions, and HIJING~\cite{Deng:2010mv} for \PbPb collisions. Efficiency corrections purely based on MC are inaccurate since the event generators do not reproduce the relative abundances of the different particle species and, in particular, they tend to significantly underestimate the production of strange hadrons. To account for this effect, a procedure based on identified hadron production measurements is followed, where the efficiency obtained from MC simulations is reweighted taking into account the primary charged particle composition measured by ALICE~\cite{ALICE:2018vuu}. This is done with data for \pp, \pPb, and \PbPb collisions at $\sqrt{s_{\rm NN}}=5.02$\,TeV~\cite{ALICE:2016dei,ALICE:2013wgn}. The residual contamination from secondary particles, i.e. particles originating from weak decays or produced in interactions with the detector material, in the selected track sample is estimated by fitting the measured ${\rm DCA}_{xy}$ distributions with a multi-component template model using as templates the DCA distributions for primary and secondary particles obtained from MC simulations~\cite{ALICE:2018vuu}.

    \begin{figure}[ht!]
    \centering
    \includegraphics[width=18pc]{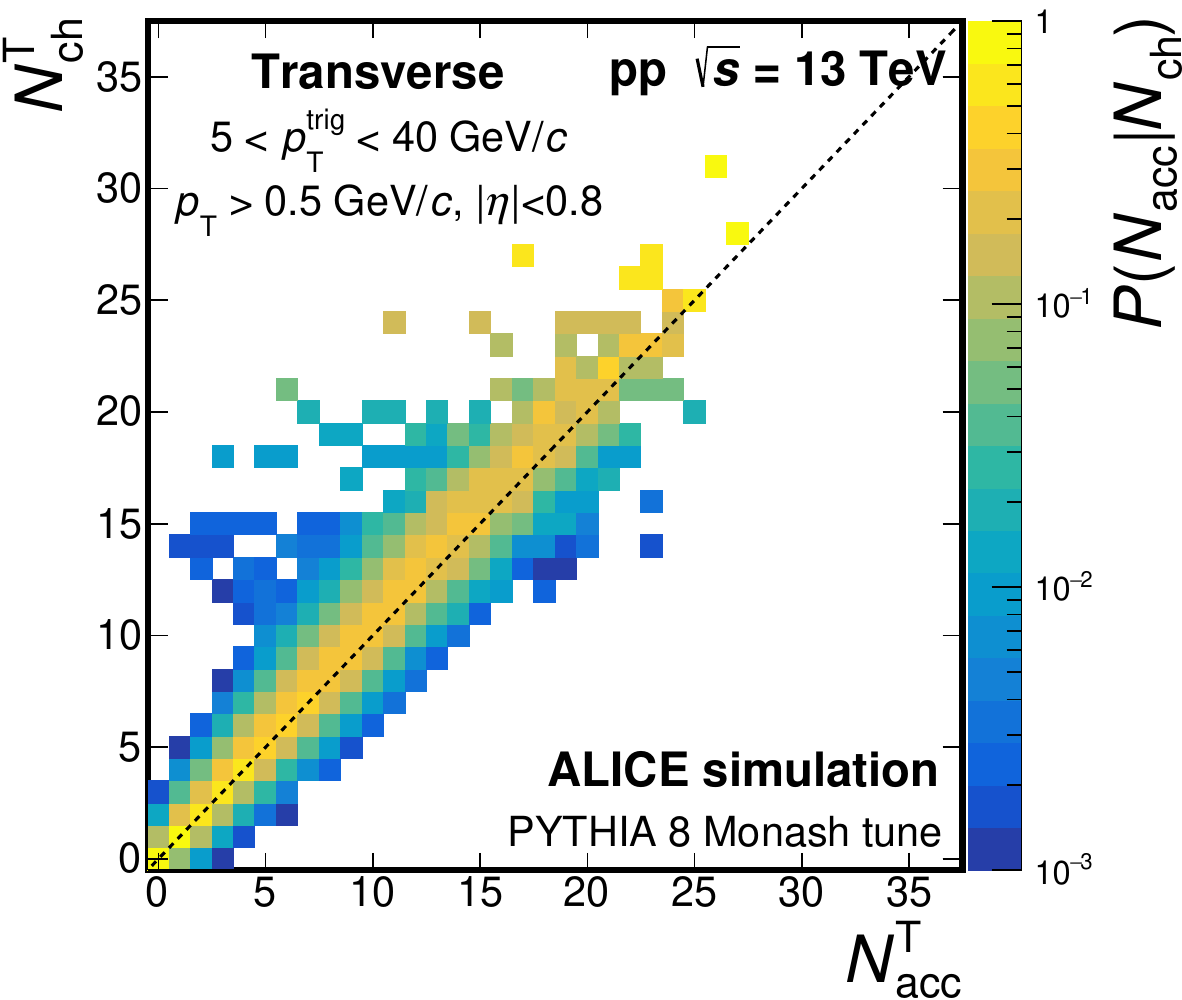}
    \hspace{5mm}
    \includegraphics[width=18pc]{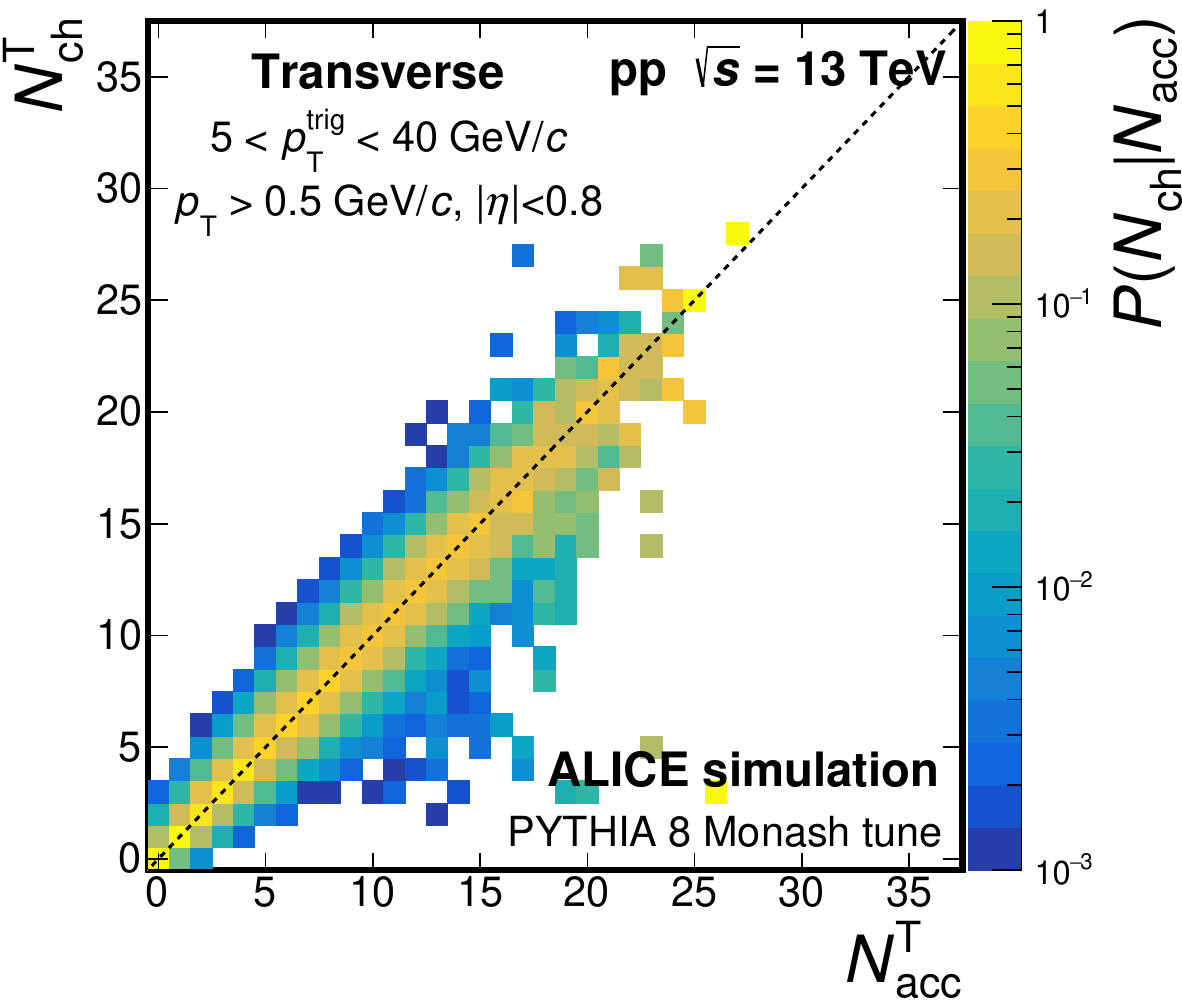}
    \caption{Response matrix $S_1$ (left) and $M_1$ matrix (right) of charged-particle multiplicity distributions in the transverse region in \pp collisions at $\s=13$\,TeV (see text for details).}
    \label{fig:RM}
    \end{figure}
    
\subsection{Bayesian unfolding of the multiplicity distributions}
    The charged-particle multiplicity in the transverse region, \nch, largely characterises the underlying-event activity. However, the measured charged-particle multiplicity distribution $Y(\nraw)$ is smeared out due to the limited acceptance and finite resolution of the detector. Hence, a one-dimensional unfolding technique based on  Bayes' theorem~\cite{dag1995}, correcting for these detector effects and efficiency losses, is introduced to recover the true multiplicity distribution. The Bayesian unfolding technique starts with the response matrix (smearing matrix) $S_1$, reflecting the detector effects on the measurements, which can be obtained from MC simulations including the transport of particles through the detector. The response matrix encodes the conditional probability $S_1 \equiv P(\nacc|\nch)$ that an event with true multiplicity \nch is measured as one with multiplicity \nacc. In the left panel of Fig.~\ref{fig:RM}, the values along the diagonal of $S_1$ represent the probability that a measured event is reconstructed with the correct charged-particle multiplicity. The off-diagonal elements give the probability that fewer (more) particles are reconstructed due to detector inefficiencies (contamination of secondaries and background particles). The one-dimensional unfolded distribution $Y(\nch)$ is given as the linear combination between the elements of the matrix $M_1$ and the measured distribution,
    \begin{equation} \label{eq:1}
        Y(\nch) = \sum_{\nraw} M_1 Y(\nraw), 
        \quad {\rm where} \quad 
        M_1 = \frac{S_1 P_0(\nch)}{\sum_{N^{\rm T}_{\rm ch}}S_1 P_0({N^{\rm T}_{\rm ch})}}.
    \end{equation} 

    $P_0(\nch)$ is a prior probability distribution, and the $M_1$ matrix represents the conditional probability $M_1 \equiv P(\nch|\nacc)$ that an event with reconstructed multiplicity \nacc has a true multiplicity \nch.  By definition, the elements of the response matrix and the $M_1$ matrix (shown in the right panel of Fig.~\ref{fig:RM}) fulfill the following normalisation conditions: $\sum_{\nacc} P(\nacc|\nch) = 1$,  $\sum_{\nch} P(\nch|\nacc) = 1$. 
    
    The unfolding technique follows an iterative process. The measured multiplicity distribution is used as the prior distribution in the first iteration. An updated prior distribution,
    \begin{equation} \label{eq:2}
        \hat{P}(\nch) = \frac{Y(\nch)}{\sum_{N^{\rm T}_{\rm ch}}Y(N^{\rm T}_{\rm ch})},
    \end{equation}
    is obtained from the second iteration onwards. Hence, the unfolding matrix is optimised as the prior distribution is updated. Finally, a new unfolded distribution can be obtained using Eq.~\ref{eq:1} with the updated $M_1$.

    After each iteration, the iterative process makes the unfolded distribution closer to the true one. Meanwhile, the statistical uncertainties in the response matrix are also propagated to the unfolded distributions through $M_1$. Thus, the uncertainties of the response matrix enter a the new unfolded distribution as $M_1$ is updated. Hence, a larger number of iterations does not guarantee a better unfolded distribution as it might be eventually contaminated by statistical fluctuations~\cite{Adye:2011gm}. In order to decide when to stop the iterations, the $\chi^2/{\rm ndf}$ between the unfolded distributions in two consecutive iterations is computed. The criterion $\chi^2/{\rm ndf}\lesssim 1$ is used to stop the iterative process.

\subsection{Unfolding of the \pt spectra as a function of \rt}
    The unfolding of the transverse momentum spectra as a function of the \rt is treated differently depending on the topological region~\cite{ALICE:2022xip,ALICE:2023yuk}. In the toward and away regions, there is no overlap between the tracks used for the spectra and the tracks used for the \rt as the multiplicity is measured in the transverse region. Hence, the one-dimensional unfolding matrix $M_1$ discussed above, can be directly applied in these two regions. In this case, the fully corrected \pt spectra as a function of \nch are obtained in a two-step procedure:
    \begin{enumerate}
    	\item The raw yields of charged particles within different \pt intervals, ${\rm d}Y(\nraw,p_{\rm T})/{\rm d}p_{\rm T}$, are reweighted using the matrix elements of $M_1$ as follows: ${\rm d}Y(\nch,p_{\rm T})/{\rm d}p_{\rm T} = \sum_{\nraw} M_1 {\rm d}Y(\nraw,p_{\rm T})/{\rm d}\pt$.  
    	\item The resulting raw yield, ${\rm d}Y(\nch,p_{\rm T})/{\rm d}p_{\rm T}$ as a function of $\nch,p_{\rm T}$, is further corrected for efficiency and secondary particle contamination.
	\end{enumerate}

    For the transverse region, a similar two-step procedure is followed. It is worth noting that in this region the multiplicity is correlated with \pt since both the spectra and multiplicity are measured using the same tracks. Thus, a specific response matrix is built based on the relationship between the \nch and \nacc distributions in each \pt interval. 
    According to the one-dimensional Bayesian unfolding technique discussed above, a corresponding  unfolding matrix $M'_1(\pt)$ is subsequently derived for each \pt interval. Hence, in the step to correct the raw yields in a given \pt interval, the $\pt$-dependent unfolding matrix $M'_1(\pt)$ is applied to the \nacc distribution as
    \begin{equation} \label{eq:3}
        Y(\nch,\pt) = \sum_{\nraw} M'_1(\pt) Y(\nraw,\pt).
    \end{equation}

\subsection{Systematic uncertainties}

    The systematic uncertainties for the \pt spectra are divided into two sets, namely the \rt-dependent and \rt-independent uncertainties.
    
    The \rt-dependent uncertainties include: 
    \begin{itemize}

    \item The systematic uncertainty due to the multiplicity dependence of the tracking efficiency is estimated from full MC simulations that include the transport of particles through the detector using GEANT~3 transport code~\cite{ALICE:2018vuu}.

     \item MC non-closure: a test is carried out by correcting the reconstructed spectra from a MC sample after full detector simulation with corrections extracted from the same generator. The MC non-closure is evaluated using the same MC production. Half of the event sample is used to obtain the correction factors and the rest to evaluate the method. The correction is expected to reproduce the input MC distribution (without detector effects) within the statistical uncertainty. The MC closure improves with the number of iterations, which is optimal at around 3. Any statistically significant difference between input and corrected distributions is referred to as MC non-closure. This difference is assigned as systematic uncertainty which is found to be 3\%, independent of \pt, for all collision systems. This contribution quantifies the accuracy of the unfolding procedure.

    \end{itemize}

    The \rt-independent uncertainties include: 
    \begin{itemize}
    
    \item Systematic uncertainty on track reconstruction and selection: this source of uncertainty is estimated by running the full analysis and varying one track selection criterion at a time~\cite{ALICE:2018vuu,ALICE:2019dfi}. The variations are similar to the ones described in Ref.~\cite{ALICE:2018vuu}. The minimum number of crossed rows in the TPC is set to 60 and 100 (the nominal is 70). The track fit quality in the ITS and TPC quantified by the $\chi^{2}_{\rm ITS}/N_{\rm hits}$ and the $\chi^{2}_{\rm TPC}/N_{\rm clusters}$ must not exceed 25 and 49 (the nominal is 36), and 3 and 5 (the nominal is 4), respectively. The maximum distance of closest approach to the vertex along the beam axis (${\rm DCA}_z$) is set to 1 and 5 cm (the nominal is 2 cm). The maximum difference between the results obtained with the tighter and looser selections with respect to the nominal values is quantified. The total systematic uncertainty on track reconstruction and selection is given as the sum in quadrature of the differences obtained from the variations of the various selections. Table~\ref{systematics} shows the systematic uncertainties due to track selection in two \pt ranges for \pp, \pPb, and \PbPb collisions.
    
    \item TPC--ITS matching efficiency: to account for the imperfect description of the probability to prolong a track from the TPC to the ITS in simulations, the track matching between the TPC--ITS track matching efficiencies in data and MC are compared after scaling the fraction of secondary particles to match the one obtained from the fits to the DCA distributions in the data. After rescaling the fraction of secondary particles, the agreement between data and MC is within 4\%~\cite{ALICE:2018vuu}. This value is assigned as an additional systematic uncertainty.

     \item Primary particle composition: the systematic uncertainty due to the reweighting of the efficiency correction is obtained from Ref.~\cite{ALICE:2018vuu}.
     
     \item Secondary particle contamination: for the systematic uncertainty from secondary particle contamination the fits to the ${\rm DCA}_{xy}$ distributions with multiple templates from the simulations is repeated using different fit intervals between 1\,cm and 3\,cm. The maximum difference between the results obtained with the different configurations with respect to the nominal value is assigned as the uncertainty due to secondary contamination.

    \item Material budget: in simulations, the density of materials used for the experimental setup was varied by $\pm$ 4.5\%~\cite{ALICE:2014sbx} to estimate the  material budget uncertainty.

     \end{itemize}
    
     As a high \ptt trigger interval ($8<\ptt<15$\,\GeVc) is chosen, the systematic uncertainty associated with event selection is negligible~\cite{ALICE:2019mmy} and the impact of the \ptt resolution is found to be negligible for the \pt spectra of associated tracks with 0.5 $<$\pt$<$ 8 GeV$/c$~\cite{ALICE:2018vuu}. The total systematic uncertainties are obtained via the sum in quadrature of the individual sources of systematic uncertainties. The ranges of systematic uncertainties from the different sources and the total systematic uncertainty are summarised in Table~\ref{systematics}.

    \begin{table}[ht!]
        \centering
        \begin{tabular}{l | c c |c c| c c}
            \hline
            \multirow{2}{4em}{Source} & \multicolumn{6}{c}{uncertainty (\%)}\\[0.5ex]
    & \multicolumn{2}{c}{\pp} & \multicolumn{2}{c}{\pPb} & \multicolumn{2}{c}{\PbPb}  \\[0.5ex] 
            \hline
            \pt (GeV$/c$)& 0.5 & 7.0 &  0.5 & 7.0 &  0.5 & 7.0  \\
            
            \hline\hline
             track reconstruction and selection* & 1.5 & 3.5 & 1.4 & 1.2 & 2.5 & 1.4  \\
             mult. dependence of tracking efficiency* & 3.0 & 3.0 & 3.0 & 3.0 &3.0 & 3.0 \\
             MC non-closure* & 3.0 & 3.0 & 3.0 & 3.0 &3.0 & 3.0   \\
             matching efficiency & 0.4 & 0.3 & 1.1 & 0.6 & 0.8 & 0.9 \\
             particle composition & 0.3& 1.3 & 0.5 & 1.2 & 0.3 & 0.7 \\
             secondary contamination &0.1 & negl. &0.3 & negl. & negl. & negl.   \\
             material budget & 0.3 & 0.2 & 0.3 & 0.2 & 0.3 & 0.2 \\
            [0.5ex]
            \hline
            Total & 4.5 & 5.7 & 4.6 & 4.6 & 5.0 & 4.6   \\
            \hline
        \end{tabular}
        \caption{Summary of systematic uncertainties for the \pt spectra of primary charged particles in the transverse region in \pp, \pPb, and \PbPb collisions at $\sqrt{s_{\rm NN}}=5.02$\,TeV. The uncertainties are shown for two representative \pt values. Sources marked with an asterisk (*) correspond to \rt-dependent systematic uncertainties.}
        \label{systematics}
    \end{table}

    For the \rt distributions in \pp collisions at different centre-of-mass energies, the systematic uncertainties include the variation of event and track selection criteria, MC non-closure uncertainties, and the uncertainties from model dependence for the unfolding. The variation of the event selection criteria is done selecting collisions within different vertex position along the $z$ axis. The maximum deviation of the results obtained by varying the vertex position (5 and 15\,cm) with respect to the result obtained using the default selection (10\,cm) is regarded as systematic uncertainty. The track selection criteria were varied for both the trigger particle and the associated particles used to measure \rt. The full simulations of the ALICE detector, including the propagation of particles through the detector, are used to get the detector response matrix. In order to evaluate the model dependence, two sets of simulations are used, and they were carried out with \py and \ep event generators, respectively. Then, two sets of corrections obtained from each MC sample are used to correct the data. The ratio of the fully corrected distribution obtained either using \py or \ep is regarded as systematic uncertainty. The model dependence is only tested for \pp collisions at $\s=13$\,TeV and the same uncertainty is assumed for lower centre-of-mass energies. A summary of the systematic uncertainties is presented in Table~\ref{tab:3}.

    \begin{table}[ht!]
    \centering
    \begin{tabular}{l |c c c| c c c| c c c| c c c}
     \hline
    \multirow{2}{4em}{Source} & \multicolumn{12}{c}{uncertainty (\%)}\\[0.5ex]
    & \multicolumn{3}{c}{2.76\,TeV} & \multicolumn{3}{c}{5.02\,TeV} & \multicolumn{3}{c}{7\,TeV} & \multicolumn{3}{c}{13\,TeV}\\[0.5ex]
    \hline\hline
    \rt & 0.5 & 2.5 & 5.0 & 0.5 & 2.5 & 5.0 & 0.5 & 2.5 & 5.0 & 0.5 & 2.5 & 5.0\\
    \hline
    event selection & 1.5 & 1.5 & 1.5 & 0.3 & 0.8 & 0.8 & 0.1 & 0.2 & 2.3 & 0.4 & 0.8 & 0.8\\
    track selection for \ptt & 2.8 & 2.8 & 2.8 & 0.4 & 1.4 & 1.4 & 0.3 & 0.5 & 2.4 & 0.2 & 1.1 & 1.2\\
    track selection for \nch & 2.7 &  7.0 & 7.0 & 0.3 & 1.0 & 4.3 & 0.4 & 2.7 & 2.7 & 0.1 & 1.1 & 5.3\\
    MC non-closure & 1.2 & 2.8 & 2.8 & 1.6 & 1.6 & 1.6 & 0.1 & 2.2 & 2.5 & 0.6 & 0.6 & 0.6\\
    model dependence & 0.9 & 1.2 & 6.5 & 0.9 & 0.6 & 7.4 & 0.9 & 0.3 & 7.4 & 0.9 & 0.8 & 7.4\\
    \hline
    Total & 4.4 & 8.3 & 10.4 & 1.9 & 2.6 & 8.9 & 1.0 & 3.5 & 8.9 & 1.2 & 2.0 & 9.2\\
    \hline
    \end{tabular}
    \caption{Summary of the systematic uncertainties for the \rt distributions at different \rt values in \pp collisions at $\s=2.76$, 5.02, 7, and 13\,TeV.}
    \label{tab:3}
    \end{table}

\section{Results and discussion}
\label{sec:results}
\subsection{Energy dependence of the relative transverse activity classifier}

\begin{figure}[ht!]
\centering
\includegraphics[width=18pc]{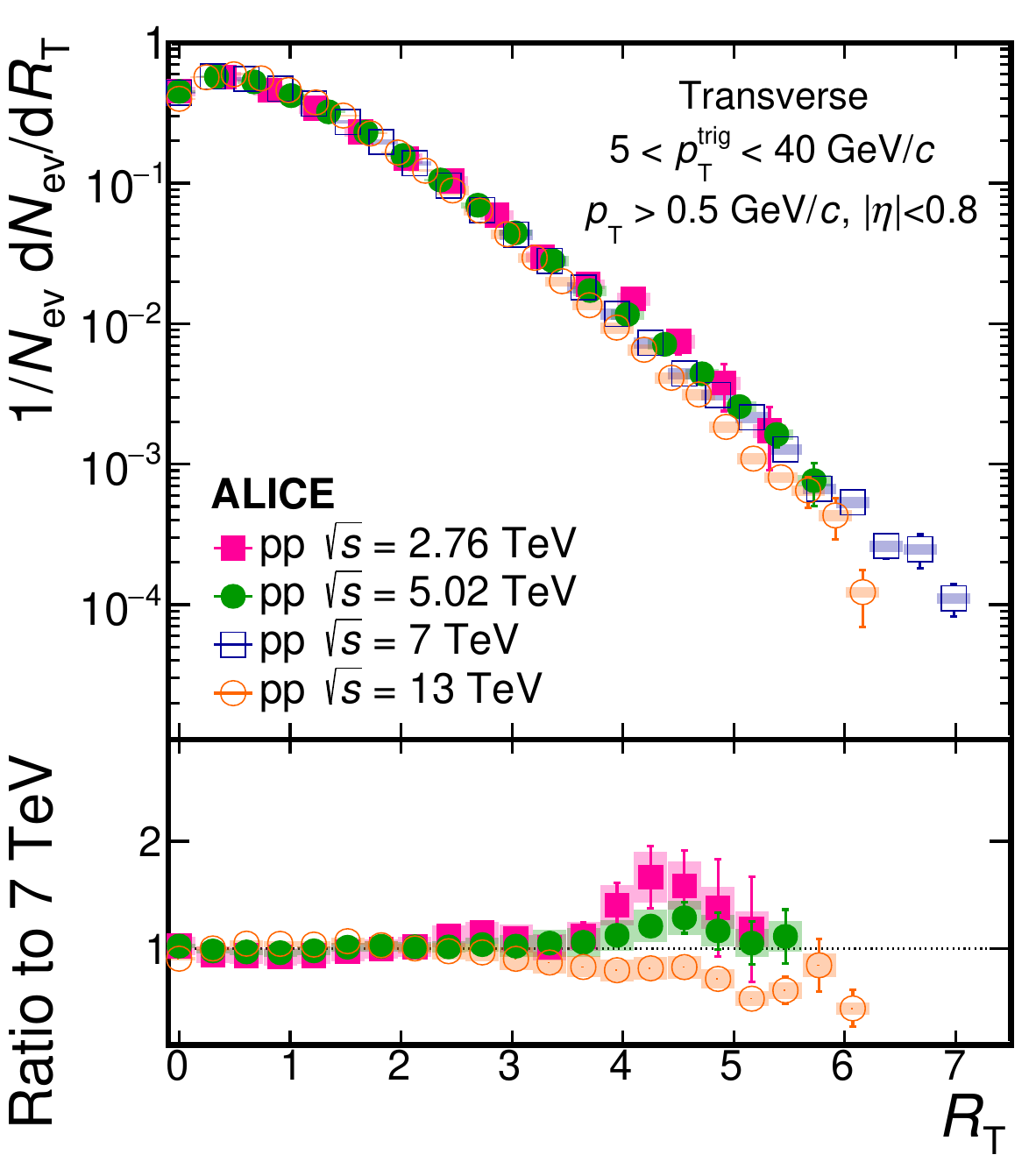}
\caption{Top: \rt distributions in \pp collisions for different centre-of-mass energies $\s=2.76$, 5.02, 7, and 13\,TeV. Bottom: \rt distributions normalised to that for \pp collisions at $\s=7$\,TeV. The ratio is calculated using a linear interpolation between adjacent points. The boxes and bars represent the systematic and statistical uncertainties, respectively. The pp sample at $\sqrt{s}=13$\,TeV is smaller than that used for pp collisions at $\sqrt{s}=7$\,TeV.}
\label{fig:kno}
\end{figure}

The measurements of the \rt distributions at different collision energies could serve as a test for the KNO scaling of particle production in the transverse region. Figure~\ref{fig:kno} shows the \rt distributions for \pp collisions at $\sqrt{s}=2.76$, 5.02, 7, and 13\,\TeV. In the bottom panel, the ratios to the \rt distribution in \pp collisions at $\s=7$\,TeV are reported.  For low \rt ($0<\rt<4$), the \rt distributions are found to be approximately (within 20\%) collision energy independent, which indicates a KNO-like scaling~\cite{Ortiz:2017jaz}. For higher \rt values ($\rt>4$), a large deviation of the ratios from unity is seen in Fig.~\ref{fig:kno}. A similar effect is observed in \py~\cite{Ortiz:2017jaz,Ortiz:2021gcr}. From an analysis aimed at measuring the MPI, it was observed that for $N_{\rm ch}/\langle N_{\rm ch}\rangle>$3--4, the number of MPI as a function of $N_{\rm ch}/\langle N_{\rm ch}\rangle$  deviates from the linear trend suggesting the presence of high-multiplicity jets~\cite{ALICE:2013tla, Ortiz:2021peu}. Since in the present analysis, high-\rt values are expected to be reached through a high number of multi-parton interactions and through a higher than average number of fragments per parton, the presence of high-multiplicity jets could explain the breaking of KNO-like scaling properties observed at high \rt in Fig.~\ref{fig:kno}. To increase the sensitivity of \rt to MPI, it has been recently proposed to build the so-called $R_{\rm T}^{\rm min}$ that is based on the charged multiplicity in the less active side of the transverse region~\cite{Bencedi:2021tst}. This approach is currently under investigation in ALICE.  

\subsection{Transverse momentum spectra as a function of the relative transverse activity classifier}

\begin{figure}[ht!]
\centering
\includegraphics[width=38pc]{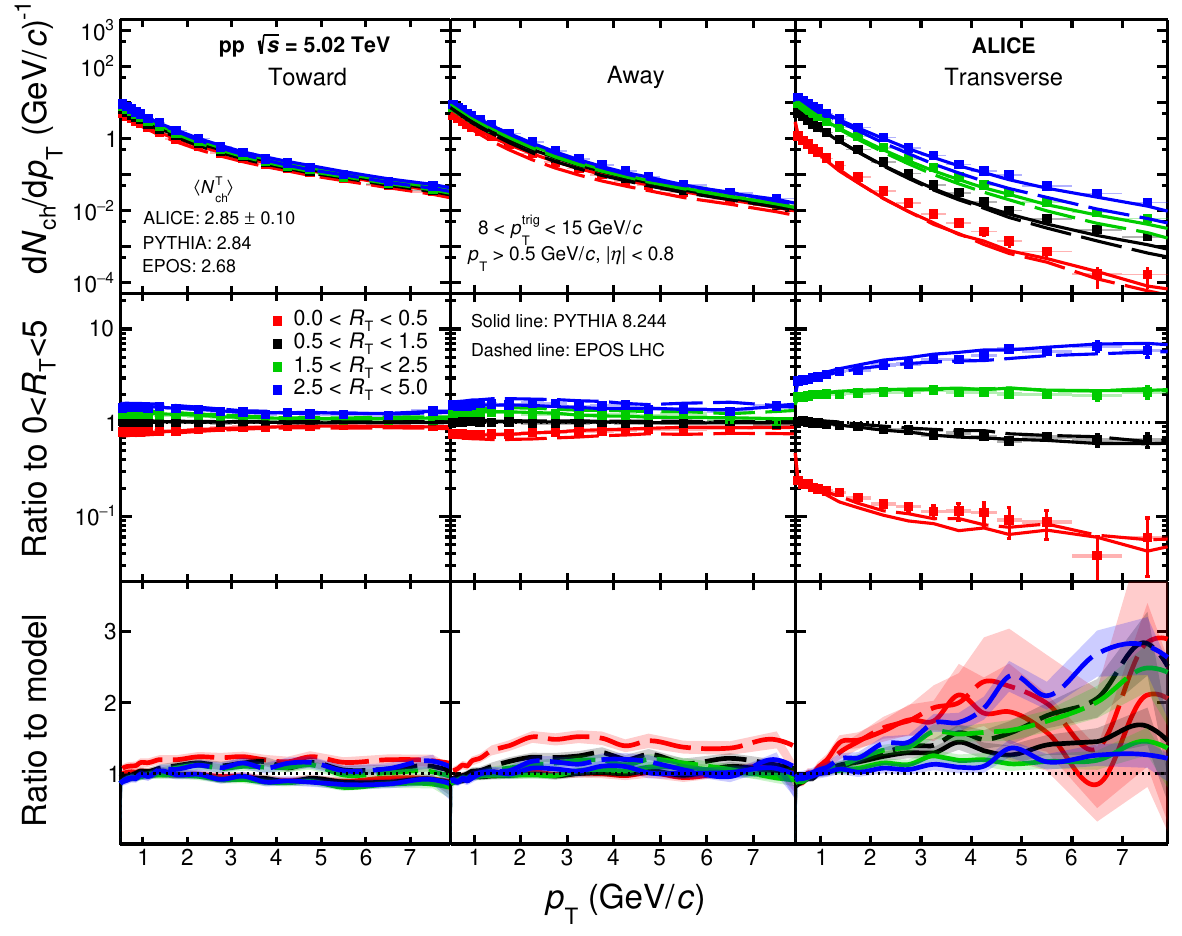}
\caption{Top panel: charged-particle transverse momentum spectra as a function of \rt for different topological regions in \pp collisions at $\sqrt{s}=5.02$\,TeV. Data are compared with \py and \ep predictions. Middle panel: the ratio of the \pt spectra in different \rt intervals to the \rt-integrated ones. The boxes and bars represent the systematic and statistical uncertainties, respectively. Bottom panel: the ratio of the \pt spectra for each \rt interval to the corresponding PYTHIA8 and EPOS-LHC predictions. The shaded area represents the sum in quadrature of the systematic and statistical uncertainties.}
\label{figmc1}
\end{figure}

\begin{figure}[ht!]
\centering
\includegraphics[width=38pc]{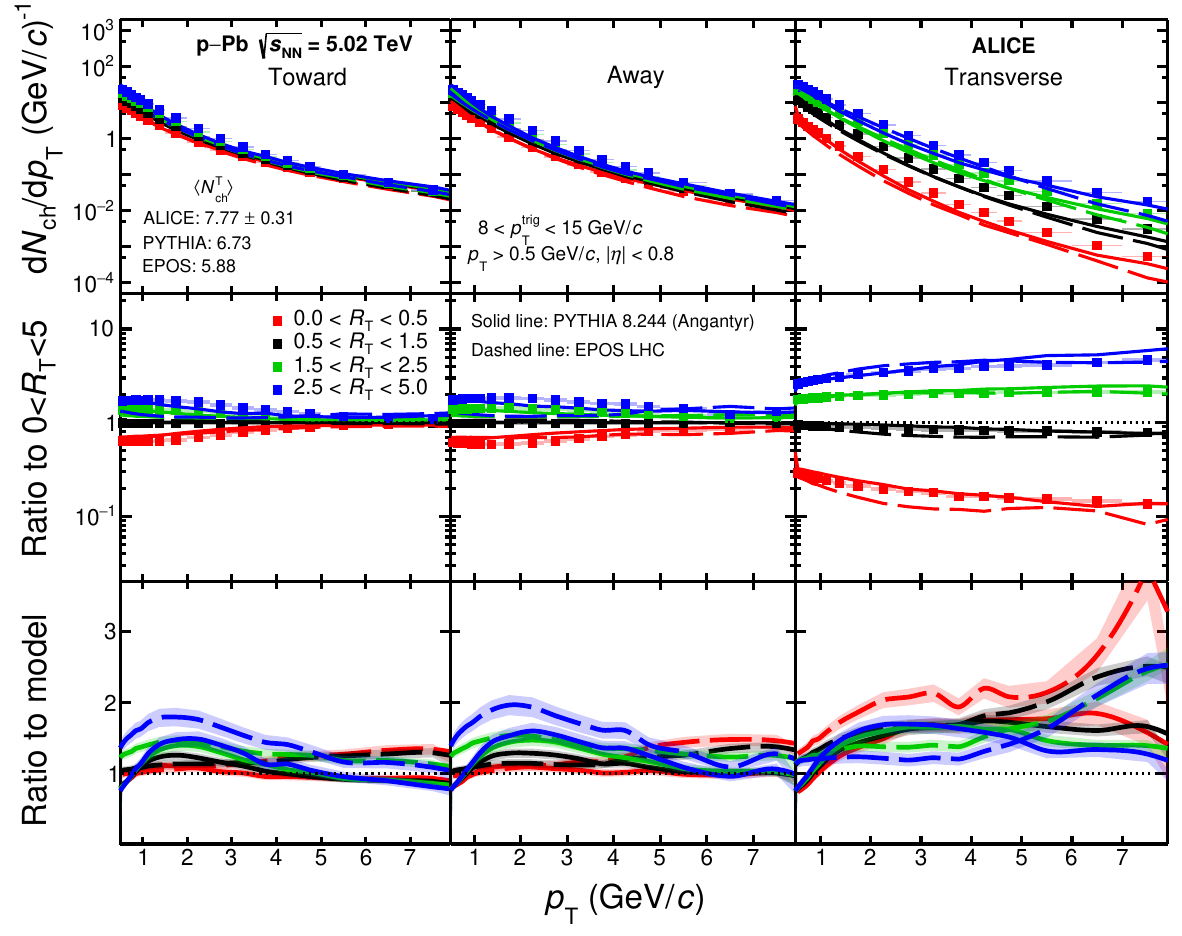}
\caption{Top panel: charged-particle transverse momentum spectra as a function of \rt for different topological regions in \pPb collisions at $\sqrt{s_{\rm NN}}=5.02$\,TeV. Data are compared with \py and \ep predictions. Middle panel: the ratio of the \pt spectra in different \rt intervals to the \rt-integrated ones. The boxes and bars represent the systematic and statistical uncertainties, respectively. Bottom panel: the ratio of the \pt spectra for each \rt interval to the corresponding PYTHIA8 Angantyr and EPOS-LHC predictions. The shaded area represents the sum in quadrature of the systematic and statistical uncertainties.}
\label{figmc2}
\end{figure}

\begin{figure}[ht!]
\centering
\includegraphics[width=38pc]{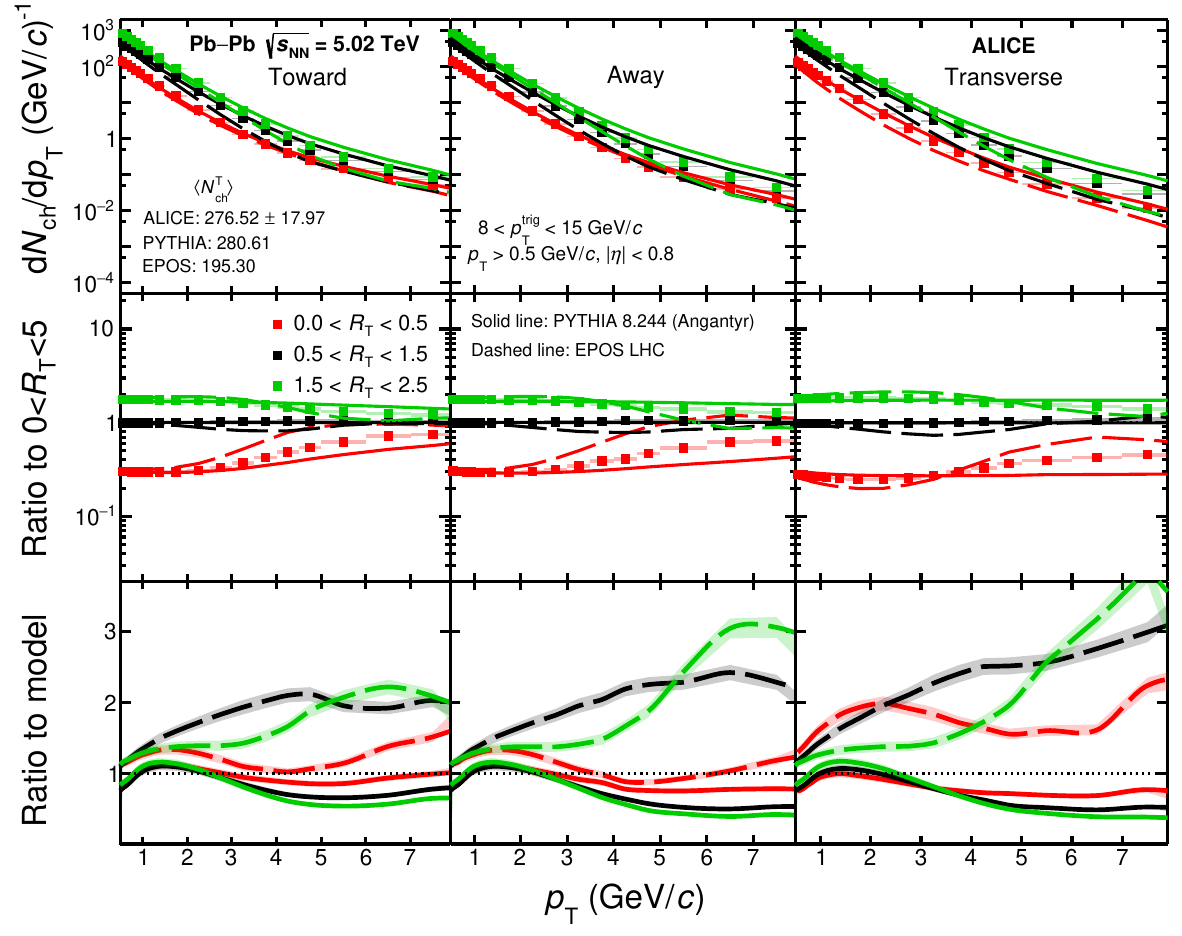}
\caption{Top panel: charged-particle transverse momentum spectra as a function of \rt for different topological regions in \pPb collisions at $\sqrt{s_{\rm NN}}=5.02$\,TeV. Data are compared with \py Angantyr and \ep predictions. Middle panel: the ratio of the \pt spectra in different \rt intervals to the \rt-integrated ones. The boxes and bars represent the systematic and statistical uncertainties, respectively. Bottom panel: the ratio of the \pt spectra for each \rt interval to the corresponding PYTHIA8 Angantyr and EPOS-LHC predictions. The shaded area represents the sum in quadrature of the systematic and statistical uncertainties.}
\label{figmc3}
\end{figure}

Figures~\ref{figmc1},~\ref{figmc2}, and~\ref{figmc3} show in the upper panels the \pt spectra of charged particles in different topological regions for different \rt intervals in \pp, \pPb, and \PbPb collisions at $\snn = 5.02$ TeV, respectively. Data are compared with predictions obtained from the event generators \py and \ep. The figures also display the average multiplicity values measured in the different collision systems. The middle panels show the ratio of the \pt spectra in different \rt intervals to the \pt spectra for the \rt-integrated sample. In general, low-\rt intervals correspond to those events with lower UE contribution and therefore are dominated by the jet fragments in the toward and away regions, while the larger values of \rt correspond to collisions with a higher UE contribution in all topological regions.

For \pp collisions, the \pt spectra in the away and toward regions relative to the \rt-integrated event class exhibit a mild \rt dependence for $\pt<4$\,GeV$/c$, where the particle yield increases with increasing \rt. This behaviour could be a trivial consequence of the amount of UE-activity in the away and toward regions, however, this effect could also be attributed to the presence of collective radial flow driving the low-\pt particles towards slightly larger momenta. This effect is mass dependent as reported in Ref.~\cite{ALICE:2023yuk}. For $\pt>4$\,GeV$/c$, the \pt spectra in the different \rt intervals converge to the \rt-integrated yield. Therefore, at higher \pt the spectral shapes in the away and toward regions are found to be almost independent of \rt. In contrast, the \pt spectra in the transverse region harden with increasing \rt. Since \rt is calculated in the transverse region, the autocorrelations are relevant in the \pt spectra measured in this region. The observed behaviour of \pt spectra in \pp collisions at $\sqrt{s}=5.02$\,TeV is similar to the results reported for \pp collisions at $\sqrt{s}=13$\,TeV in Ref.~\cite{ALICE:2023yuk}. For \pPb collisions the data show compatible features with those of \pp collisions in all the three topological regions.

The behaviour of the charged-particle \pt spectra as a function of \rt in the toward and away regions in \PbPb collisions, where the jet bias is nearly absent in the transverse region, is qualitatively similar to that of \pp and \pPb collisions. This is understood with the fact that in all three collision systems, the production of particles in the toward and away regions is dominated by the fragmentation of the two outgoing hard partons and hence dominated by particles with high \pt.
However, the \pt spectra in \rt intervals converge to the \rt-integrated result at higher \pt values ($\pt>~6$ GeV$/c$) as compared to \pp and \pPb collisions. The behaviour in the transverse region the \pt spectra in \PbPb collisions is found to be similar to that in toward and away regions.

In Figs.~\ref{figmc1},~\ref{figmc2}, and~\ref{figmc3}, the measured spectra are also compared to \py and \ep predictions and the bottom panels show data to model ratios. For \pp collisions, \py describes the \rt dependence of the transverse momentum spectra in the full \pt interval for the toward and away regions. A similar level of agreement is observed when comparing \ep with data except for $R_{\rm T}<0.5$ in the away region. For the transverse region, \py captures the spectral shapes within two sigmas while \ep deviates from data, in particular at high \pt.

For \pPb collisions, \py Angantyr describes the high \pt yield ($p_{\rm T}>4$\,GeV$/c$) for the toward and away regions better than \ep. The models are able to describe the yield for $R_{\rm T}<1.5$, but they underestimate the data,especially at low \pt, in the highest \rt interval. For the transverse region, none of the used models is able to capture the \rt dependence of the \pt spectra.

For \PbPb collisions, \py Angantyr overestimates the high \pt yield ($p_{\rm T}>3$ GeV$/c$) for all three topological regions, but it fairly describes the data in the lower \pt region. In contrast, \ep underestimates the yields in the three topological regions and the deviations from data increase with increasing \pt.

For the ratios of the \pt spectra in different \rt intervals to the inclusive spectra, the two models agree with each other in \pp and \pPb collisions, and also they agree with the experimental data. For Pb–Pb collisions, \ep gives a better qualitative description of the \pt-dependent trend observed in data compared with \py Angantyr for \pt $>$ 4 \GeVc. \py Angantyr tends to give a good description only for $\pt<4$\,GeV$/c$.

Note that the measured value of $\langle \nch \rangle$ in pp collisions is found to be consistent with the \py prediction while \ep deviates from the measured value by 6\%. For \pPb collisions, the values of $\langle \nch \rangle$ deviate from data by 13\% and 24\% for \py Angantyr and \ep, respectively. Similarly for \PbPb collisions, the values of $\langle \nch \rangle$ deviate from data by 1.5\% and 29\% for \py Angantyr and \ep, respectively.

\subsection{Average transverse momentum and integrated yield as a function of the relative transverse activity classifier}

\begin{figure}[ht!]
\centering
\includegraphics[width=38pc]{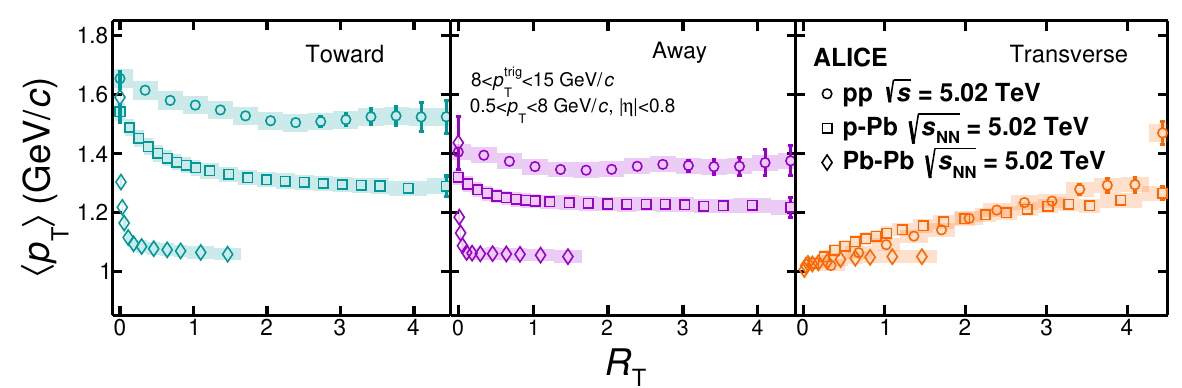}
\caption{\mpt of charged particles in toward (left), away (middle), and transverse (right) regions as a function of \rt for \pp, \pPb, and \PbPb collisions at $\snn=5.02$\,TeV.}
\label{fig12}
\end{figure}

Figure~\ref{fig12} shows the \rt dependence of the mean transverse momentum, \mpt, derived from the \pt spectra of charged particles in the measured \pt range. The systematic uncertainties assigned to the \pt spectra were propagated to \mpt~\cite{ALICE:2013rdo}. Results are shown for the three topological regions as a function of \rt for \pp, \pPb, and \PbPb collisions at $\snn= 5.02$\,TeV. The increasing trend of \mpt in the transverse region with increasing \rt for the three collision system is similar to the results reported in Ref.~\cite{ALICE:2013rdo}. For the toward and away regions, the \mpt seems to be nearly flat as a function of \rt, except for $\rt < 1$, where the \mpt increases with decreasing \rt. For $\rt > 1$, the \mpt values in the toward and away regions are higher in \pp than in \pPb (and \pPb is higher than \PbPb) due to the larger contribution from the UE particles (which are softer than those of the leading jets) for larger collision systems.

\begin{figure}[ht!]
\centering
\includegraphics[width=38pc]{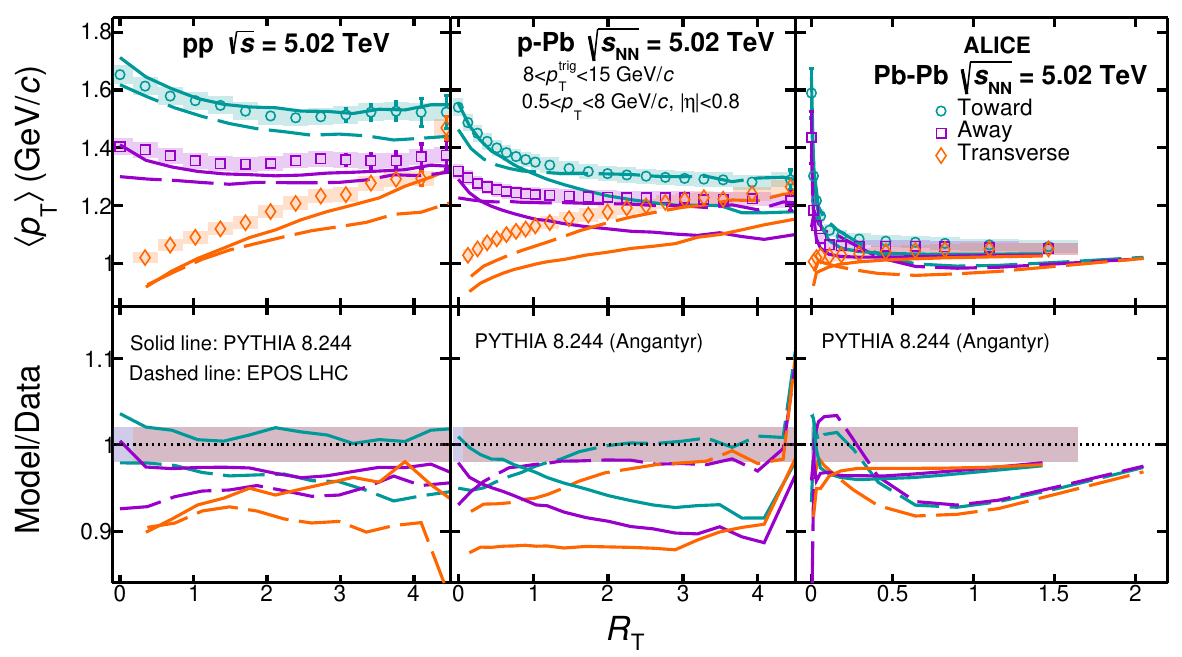}
\caption{Top: \mpt of charged particles in \pp (left), \pPb (middle), and \PbPb (right) collisions at $\snn=5.02$\,TeV as a function of \rt for different topological regions compared with predictions from \ep and \py. Bottom: ratio of MC to data. The band around unity in the ratio depicts the experimental uncertainties.}
\label{fig12.001}
\end{figure}

At $\rt \approx 0$, the \mpt is observed to be similar in the three collision systems. Such behaviour is expected since the contribution of jets dominates at low-\rt and hence all collision systems are expected to approach the pp collisions limit.
At high values of \rt, where the UE contribution is dominant, the \mpt values are similar in all three topological regions for a given collision system. The \mpt is compared with the predictions from \ep and \py in Fig.~\ref{fig12.001}. The models deviate by 10--20\% from the experimental data, however, they show a trend with \rt that is qualitatively similar to the measured one.

\begin{figure}[ht!]
\centering
\includegraphics[width=38pc]{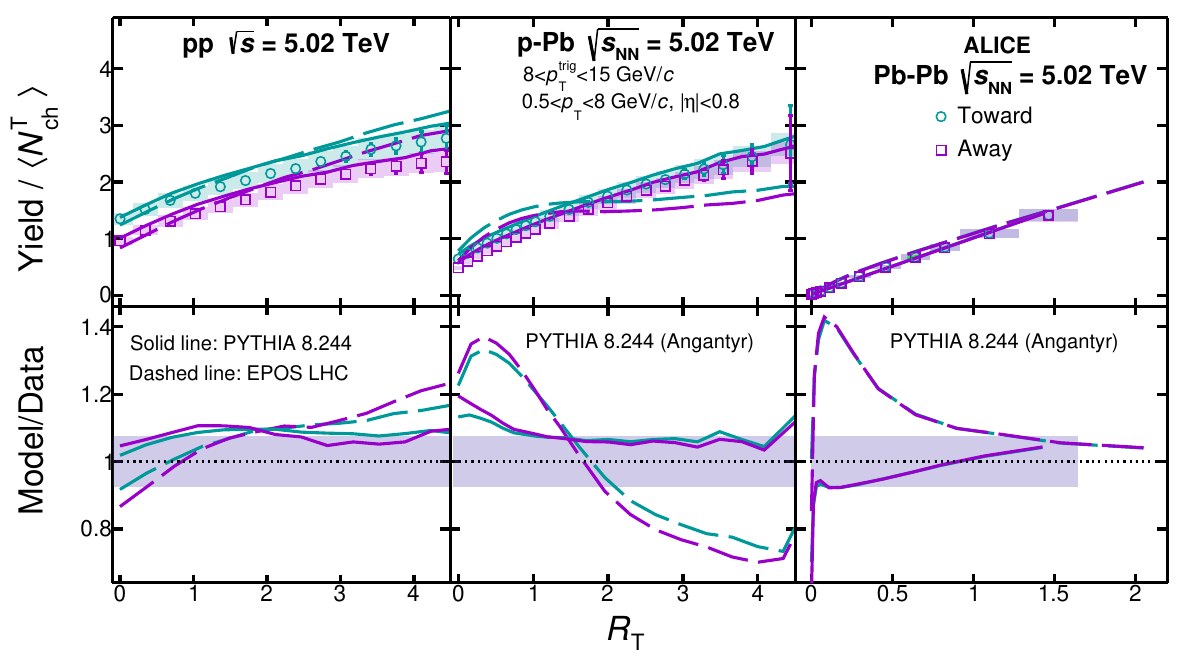}
\caption{Top: integrated yield of charged particles in \pp (left), \pPb (middle), and \PbPb (right) collisions at $\snn=5.02$\,TeV as a function of \rt for toward and away regions compared with predictions from \ep and \py. Bottom: ratio of model predictions to data. The band around unity in the ratio depicts the experimental uncertainties.}
\label{fig11}
\end{figure}

Finally, the particle yields in the toward and away regions relative to those in the transverse region are studied as a function of \rt. In pp collisions, due to the limited ALICE acceptance the particle multiplicities in the toward and away regions are expected to be proportional to that in the transverse region only for $R_{\rm T}<2$. For example, it has been shown that the MPI activity (UE) is proportional to \rt only up to $R_{\rm T}=2$~\cite{Bencedi:2020qpi}. At higher \rt, the sample is biased towards multi-jet final states producing a hardening of the \pt spectra in the transverse region. In order to further investigate this bias, Fig.~\ref{fig11} shows the \pt-integrated yield normalised to $\langle \nch \rangle$ of charged particles in \pp (left), \pPb (middle), and \PbPb (right) collisions as a function of \rt for the toward and away regions at $\snn=5.02$\,TeV. The \nch-normalised integrated yield is also compared with the predictions from \ep and \py. The lower panels show the ratio of the model predictions to the data. For \pp and \pPb collisions, the normalised integrated yields for both toward and away regions show a linear increase for $R_{\rm T}<2$ and then they tend to saturate at high \rt. \py describes qualitatively the trend of the data while \ep shows no hint of saturation and thus overestimates the data at high \rt in \pp collisions, while in \pPb collisions it predicts a completely different trend compared to the measured one. For \PbPb collisions, the behaviour of the normalised yield in both models seems to follow a linear trend. Both models show a different trend at low-\rt, however, the prediction from \py is found to be quantitatively consistent with the data within uncertainties. The \ep prediction shows about 40\% higher yield with respect to the data at low-\rt and approaches the measured values at high \rt.

\section{Conclusion}
\label{sec:conclusion}

The \rt distributions have been measured in \pp collisions at $\s=2.76$, 5.02, 7, and 13\,TeV. They exhibit a Koba-Nielsen-Olesen (KNO)-like scaling for events with low-\rt, while for UE-dominated events characterised by high \rt, an indication of violation of the scaling is seen which might be attributed to high-multiplicity jets.  In order to investigate the UE properties, the \pt spectra as a function of \rt have been studied at $\snn=5.02$\,TeV in different collision systems. In general, the  charged-particle spectra in \rt intervals show a similar behaviour in \pp and \pPb collisions for all the three topological regions (toward, away, and transverse). The transverse region exhibits autocorrelation effects that give rise to a hardening of the spectra with increasing \pt, in contrast to the toward and away regions, where particle production at low-\pt increases (decreases) with increasing (decreasing) \rt. At higher \pt the spectral shapes become almost independent of \rt. In these collision systems, a softening of the spectra occurs with increasing \rt. In contrast to the small systems, \PbPb collisions do not show an autocorrelation effect in the transverse region. In this case, the behaviour of the three topological regions seems to be dominated by soft interactions. 

Subsequently, the study of the \mpt of charged particles in the three topological regions as a function of \rt for the three collision systems shows that at $\rt \sim 0$, the \mpt is, within the current uncertainties, independent of collision system size and such a limit is well described by models. Overall, \py gives a better description of the data presented in this paper than \ep. The experimental results presented in this article will provide important contributions towards the development and tuning of MC event generators. 


\newenvironment{acknowledgement}{\relax}{\relax}
\begin{acknowledgement}
\section*{Acknowledgements}


The ALICE Collaboration would like to thank all its engineers and technicians for their invaluable contributions to the construction of the experiment and the CERN accelerator teams for the outstanding performance of the LHC complex.
The ALICE Collaboration gratefully acknowledges the resources and support provided by all Grid centres and the Worldwide LHC Computing Grid (WLCG) collaboration.
The ALICE Collaboration acknowledges the following funding agencies for their support in building and running the ALICE detector:
A. I. Alikhanyan National Science Laboratory (Yerevan Physics Institute) Foundation (ANSL), State Committee of Science and World Federation of Scientists (WFS), Armenia;
Austrian Academy of Sciences, Austrian Science Fund (FWF): [M 2467-N36] and Nationalstiftung f\"{u}r Forschung, Technologie und Entwicklung, Austria;
Ministry of Communications and High Technologies, National Nuclear Research Center, Azerbaijan;
Conselho Nacional de Desenvolvimento Cient\'{\i}fico e Tecnol\'{o}gico (CNPq), Financiadora de Estudos e Projetos (Finep), Funda\c{c}\~{a}o de Amparo \`{a} Pesquisa do Estado de S\~{a}o Paulo (FAPESP) and Universidade Federal do Rio Grande do Sul (UFRGS), Brazil;
Bulgarian Ministry of Education and Science, within the National Roadmap for Research Infrastructures 2020-2027 (object CERN), Bulgaria;
Ministry of Education of China (MOEC) , Ministry of Science \& Technology of China (MSTC) and National Natural Science Foundation of China (NSFC), China;
Ministry of Science and Education and Croatian Science Foundation, Croatia;
Centro de Aplicaciones Tecnol\'{o}gicas y Desarrollo Nuclear (CEADEN), Cubaenerg\'{\i}a, Cuba;
Ministry of Education, Youth and Sports of the Czech Republic, Czech Republic;
The Danish Council for Independent Research | Natural Sciences, the VILLUM FONDEN and Danish National Research Foundation (DNRF), Denmark;
Helsinki Institute of Physics (HIP), Finland;
Commissariat \`{a} l'Energie Atomique (CEA) and Institut National de Physique Nucl\'{e}aire et de Physique des Particules (IN2P3) and Centre National de la Recherche Scientifique (CNRS), France;
Bundesministerium f\"{u}r Bildung und Forschung (BMBF) and GSI Helmholtzzentrum f\"{u}r Schwerionenforschung GmbH, Germany;
General Secretariat for Research and Technology, Ministry of Education, Research and Religions, Greece;
National Research, Development and Innovation Office, Hungary;
Department of Atomic Energy Government of India (DAE), Department of Science and Technology, Government of India (DST), University Grants Commission, Government of India (UGC) and Council of Scientific and Industrial Research (CSIR), India;
National Research and Innovation Agency - BRIN, Indonesia;
Istituto Nazionale di Fisica Nucleare (INFN), Italy;
Japanese Ministry of Education, Culture, Sports, Science and Technology (MEXT) and Japan Society for the Promotion of Science (JSPS) KAKENHI, Japan;
Consejo Nacional de Ciencia (CONACYT) y Tecnolog\'{i}a, through Fondo de Cooperaci\'{o}n Internacional en Ciencia y Tecnolog\'{i}a (FONCICYT) and Direcci\'{o}n General de Asuntos del Personal Academico (DGAPA), Mexico;
Nederlandse Organisatie voor Wetenschappelijk Onderzoek (NWO), Netherlands;
The Research Council of Norway, Norway;
Commission on Science and Technology for Sustainable Development in the South (COMSATS), Pakistan;
Pontificia Universidad Cat\'{o}lica del Per\'{u}, Peru;
Ministry of Education and Science, National Science Centre and WUT ID-UB, Poland;
Korea Institute of Science and Technology Information and National Research Foundation of Korea (NRF), Republic of Korea;
Ministry of Education and Scientific Research, Institute of Atomic Physics, Ministry of Research and Innovation and Institute of Atomic Physics and Universitatea Nationala de Stiinta si Tehnologie Politehnica Bucuresti, Romania;
Ministry of Education, Science, Research and Sport of the Slovak Republic, Slovakia;
National Research Foundation of South Africa, South Africa;
Swedish Research Council (VR) and Knut \& Alice Wallenberg Foundation (KAW), Sweden;
European Organization for Nuclear Research, Switzerland;
Suranaree University of Technology (SUT), National Science and Technology Development Agency (NSTDA) and National Science, Research and Innovation Fund (NSRF via PMU-B B05F650021), Thailand;
Turkish Energy, Nuclear and Mineral Research Agency (TENMAK), Turkey;
National Academy of  Sciences of Ukraine, Ukraine;
Science and Technology Facilities Council (STFC), United Kingdom;
National Science Foundation of the United States of America (NSF) and United States Department of Energy, Office of Nuclear Physics (DOE NP), United States of America.
In addition, individual groups or members have received support from:
European Research Council, Strong 2020 - Horizon 2020 (grant nos. 950692, 824093), European Union;
Academy of Finland (Center of Excellence in Quark Matter) (grant nos. 346327, 346328), Finland.
\end{acknowledgement}

\bibliographystyle{utphys}   
\bibliography{bibliography}

\newpage
\appendix

%
%

\section{The ALICE Collaboration}
\label{app:collab}
\begin{flushleft} 
\small

S.~Acharya\,\orcidlink{0000-0002-9213-5329}\,$^{\rm 128}$, 
D.~Adamov\'{a}\,\orcidlink{0000-0002-0504-7428}\,$^{\rm 87}$, 
G.~Aglieri Rinella\,\orcidlink{0000-0002-9611-3696}\,$^{\rm 33}$, 
M.~Agnello\,\orcidlink{0000-0002-0760-5075}\,$^{\rm 30}$, 
N.~Agrawal\,\orcidlink{0000-0003-0348-9836}\,$^{\rm 52}$, 
Z.~Ahammed\,\orcidlink{0000-0001-5241-7412}\,$^{\rm 136}$, 
S.~Ahmad\,\orcidlink{0000-0003-0497-5705}\,$^{\rm 16}$, 
S.U.~Ahn\,\orcidlink{0000-0001-8847-489X}\,$^{\rm 72}$, 
I.~Ahuja\,\orcidlink{0000-0002-4417-1392}\,$^{\rm 38}$, 
A.~Akindinov\,\orcidlink{0000-0002-7388-3022}\,$^{\rm 142}$
% ...

\section*{Affiliation Notes}

$^{\rm I}$ Deceased\\
$^{\rm II}$ Also at: Max-Planck-Institut fur Physik, Munich, Germany\\
$^{\rm III}$ Also at: Italian National Agency for New Technologies, Energy and Sustainable Economic Development (ENEA), Bologna, Italy\\
$^{\rm IV}$ Also at: Dipartimento DET del Politecnico di Torino, Turin, Italy\\
$^{\rm V}$ Also at: Department of Applied Physics, Aligarh Muslim University, Aligarh, India\\
$^{\rm VI}$ Also at: Institute of Theoretical Physics, University of Wroclaw, Poland\\
$^{\rm VII}$ Also at: An institution covered by a cooperation agreement with CERN\\

\section*{Collaboration Institutes}

$^{1}$ A.I. Alikhanyan National Science Laboratory (Yerevan Physics Institute) Foundation, Yerevan, Armenia\\
$^{2}$ AGH University of Krakow, Cracow, Poland\\
$^{3}$ Bogolyubov Institute for Theoretical Physics, National Academy of Sciences of Ukraine, Kiev, Ukraine\\
$^{4}$ Bose Institute, Department of Physics  and Centre for Astroparticle Physics and Space Science (CAPSS), Kolkata, India\\
$^{5}$ California Polytechnic State University, San Luis Obispo, California, United States\\
$^{6}$ Central China Normal University, Wuhan, China\\
$^{7}$ Centro de Aplicaciones Tecnol\'{o}gicas y Desarrollo Nuclear (CEADEN), Havana, Cuba\\
$^{8}$ Centro de Investigaci\'{o}n y de Estudios Avanzados (CINVESTAV), Mexico City and M\'{e}rida, Mexico\\
$^{9}$ Chicago State University, Chicago, Illinois, United States\\
$^{10}$ China Institute of Atomic Energy, Beijing, China\\
$^{11}$ China University of Geosciences, Wuhan, China\\
$^{12}$ Chungbuk National University, Cheongju, Republic of Korea\\
$^{13}$ Comenius University Bratislava, Faculty of Mathematics, Physics and Informatics, Bratislava, Slovak Republic\\
$^{14}$ COMSATS University Islamabad, Islamabad, Pakistan\\
$^{15}$ Creighton University, Omaha, Nebraska, United States\\
$^{16}$ Department of Physics, Aligarh Muslim University, Aligarh, India\\
$^{17}$ Department of Physics, Pusan National University, Pusan, Republic of Korea\\
$^{18}$ Department of Physics, Sejong University, Seoul, Republic of Korea\\
$^{19}$ Department of Physics, University of California, Berkeley, California, United States\\
$^{20}$ Department of Physics, University of Oslo, Oslo, Norway\\
$^{21}$ Department of Physics and Technology, University of Bergen, Bergen, Norway\\
$^{22}$ Dipartimento di Fisica, Universit\`{a} di Pavia, Pavia, Italy\\
$^{23}$ Dipartimento di Fisica dell'Universit\`{a} and Sezione INFN, Cagliari, Italy\\
$^{24}$ Dipartimento di Fisica dell'Universit\`{a} and Sezione INFN, Trieste, Italy\\
$^{25}$ Dipartimento di Fisica dell'Universit\`{a} and Sezione INFN, Turin, Italy\\
$^{26}$ Dipartimento di Fisica e Astronomia dell'Universit\`{a} and Sezione INFN, Bologna, Italy\\
$^{27}$ Dipartimento di Fisica e Astronomia dell'Universit\`{a} and Sezione INFN, Catania, Italy\\
$^{28}$ Dipartimento di Fisica e Astronomia dell'Universit\`{a} and Sezione INFN, Padova, Italy\\
$^{29}$ Dipartimento di Fisica `E.R.~Caianiello' dell'Universit\`{a} and Gruppo Collegato INFN, Salerno, Italy\\
$^{30}$ Dipartimento DISAT del Politecnico and Sezione INFN, Turin, Italy\\
$^{31}$ Dipartimento di Scienze MIFT, Universit\`{a} di Messina, Messina, Italy\\
$^{32}$ Dipartimento Interateneo di Fisica `M.~Merlin' and Sezione INFN, Bari, Italy\\
$^{33}$ European Organization for Nuclear Research (CERN), Geneva, Switzerland\\
$^{34}$ Faculty of Electrical Engineering, Mechanical Engineering and Naval Architecture, University of Split, Split, Croatia\\
$^{35}$ Faculty of Engineering and Science, Western Norway University of Applied Sciences, Bergen, Norway\\
$^{36}$ Faculty of Nuclear Sciences and Physical Engineering, Czech Technical University in Prague, Prague, Czech Republic\\
$^{37}$ Faculty of Physics, Sofia University, Sofia, Bulgaria\\
$^{38}$ Faculty of Science, P.J.~\v{S}af\'{a}rik University, Ko\v{s}ice, Slovak Republic\\
$^{39}$ Frankfurt Institute for Advanced Studies, Johann Wolfgang Goethe-Universit\"{a}t Frankfurt, Frankfurt, Germany\\
$^{40}$ Fudan University, Shanghai, China\\
$^{41}$ Gangneung-Wonju National University, Gangneung, Republic of Korea\\
$^{42}$ Gauhati University, Department of Physics, Guwahati, India\\
$^{43}$ Helmholtz-Institut f\"{u}r Strahlen- und Kernphysik, Rheinische Friedrich-Wilhelms-Universit\"{a}t Bonn, Bonn, Germany\\
$^{44}$ Helsinki Institute of Physics (HIP), Helsinki, Finland\\
$^{45}$ High Energy Physics Group,  Universidad Aut\'{o}noma de Puebla, Puebla, Mexico\\
$^{46}$ Horia Hulubei National Institute of Physics and Nuclear Engineering, Bucharest, Romania\\
$^{47}$ HUN-REN Wigner Research Centre for Physics, Budapest, Hungary\\
$^{48}$ Indian Institute of Technology Bombay (IIT), Mumbai, India\\
$^{49}$ Indian Institute of Technology Indore, Indore, India\\
$^{50}$ INFN, Laboratori Nazionali di Frascati, Frascati, Italy\\
$^{51}$ INFN, Sezione di Bari, Bari, Italy\\
$^{52}$ INFN, Sezione di Bologna, Bologna, Italy\\
$^{53}$ INFN, Sezione di Cagliari, Cagliari, Italy\\
$^{54}$ INFN, Sezione di Catania, Catania, Italy\\
$^{55}$ INFN, Sezione di Padova, Padova, Italy\\
$^{56}$ INFN, Sezione di Pavia, Pavia, Italy\\
$^{57}$ INFN, Sezione di Torino, Turin, Italy\\
$^{58}$ INFN, Sezione di Trieste, Trieste, Italy\\
$^{59}$ Inha University, Incheon, Republic of Korea\\
$^{60}$ Institute for Gravitational and Subatomic Physics (GRASP), Utrecht University/Nikhef, Utrecht, Netherlands\\
$^{61}$ Institute of Experimental Physics, Slovak Academy of Sciences, Ko\v{s}ice, Slovak Republic\\
$^{62}$ Institute of Physics, Homi Bhabha National Institute, Bhubaneswar, India\\
$^{63}$ Institute of Physics of the Czech Academy of Sciences, Prague, Czech Republic\\
$^{64}$ Institute of Space Science (ISS), Bucharest, Romania\\
$^{65}$ Institut f\"{u}r Kernphysik, Johann Wolfgang Goethe-Universit\"{a}t Frankfurt, Frankfurt, Germany\\
$^{66}$ Instituto de Ciencias Nucleares, Universidad Nacional Aut\'{o}noma de M\'{e}xico, Mexico City, Mexico\\
$^{67}$ Instituto de F\'{i}sica, Universidade Federal do Rio Grande do Sul (UFRGS), Porto Alegre, Brazil\\
$^{68}$ Instituto de F\'{\i}sica, Universidad Nacional Aut\'{o}noma de M\'{e}xico, Mexico City, Mexico\\
$^{69}$ iThemba LABS, National Research Foundation, Somerset West, South Africa\\
$^{70}$ Jeonbuk National University, Jeonju, Republic of Korea\\
$^{71}$ Johann-Wolfgang-Goethe Universit\"{a}t Frankfurt Institut f\"{u}r Informatik, Fachbereich Informatik und Mathematik, Frankfurt, Germany\\
$^{72}$ Korea Institute of Science and Technology Information, Daejeon, Republic of Korea\\
$^{73}$ KTO Karatay University, Konya, Turkey\\
$^{74}$ Laboratoire de Physique Subatomique et de Cosmologie, Universit\'{e} Grenoble-Alpes, CNRS-IN2P3, Grenoble, France\\
$^{75}$ Lawrence Berkeley National Laboratory, Berkeley, California, United States\\
$^{76}$ Lund University Department of Physics, Division of Particle Physics, Lund, Sweden\\
$^{77}$ Nagasaki Institute of Applied Science, Nagasaki, Japan\\
$^{78}$ Nara Women{'}s University (NWU), Nara, Japan\\
$^{79}$ National and Kapodistrian University of Athens, School of Science, Department of Physics , Athens, Greece\\
$^{80}$ National Centre for Nuclear Research, Warsaw, Poland\\
$^{81}$ National Institute of Science Education and Research, Homi Bhabha National Institute, Jatni, India\\
$^{82}$ National Nuclear Research Center, Baku, Azerbaijan\\
$^{83}$ National Research and Innovation Agency - BRIN, Jakarta, Indonesia\\
$^{84}$ Niels Bohr Institute, University of Copenhagen, Copenhagen, Denmark\\
$^{85}$ Nikhef, National institute for subatomic physics, Amsterdam, Netherlands\\
$^{86}$ Nuclear Physics Group, STFC Daresbury Laboratory, Daresbury, United Kingdom\\
$^{87}$ Nuclear Physics Institute of the Czech Academy of Sciences, Husinec-\v{R}e\v{z}, Czech Republic\\
$^{88}$ Oak Ridge National Laboratory, Oak Ridge, Tennessee, United States\\
$^{89}$ Ohio State University, Columbus, Ohio, United States\\
$^{90}$ Physics department, Faculty of science, University of Zagreb, Zagreb, Croatia\\
$^{91}$ Physics Department, Panjab University, Chandigarh, India\\
$^{92}$ Physics Department, University of Jammu, Jammu, India\\
$^{93}$ Physics Program and International Institute for Sustainability with Knotted Chiral Meta Matter (SKCM2), Hiroshima University, Hiroshima, Japan\\
$^{94}$ Physikalisches Institut, Eberhard-Karls-Universit\"{a}t T\"{u}bingen, T\"{u}bingen, Germany\\
$^{95}$ Physikalisches Institut, Ruprecht-Karls-Universit\"{a}t Heidelberg, Heidelberg, Germany\\
$^{96}$ Physik Department, Technische Universit\"{a}t M\"{u}nchen, Munich, Germany\\
$^{97}$ Politecnico di Bari and Sezione INFN, Bari, Italy\\
$^{98}$ Research Division and ExtreMe Matter Institute EMMI, GSI Helmholtzzentrum f\"ur Schwerionenforschung GmbH, Darmstadt, Germany\\
$^{99}$ Saga University, Saga, Japan\\
$^{100}$ Saha Institute of Nuclear Physics, Homi Bhabha National Institute, Kolkata, India\\
$^{101}$ School of Physics and Astronomy, University of Birmingham, Birmingham, United Kingdom\\
$^{102}$ Secci\'{o}n F\'{\i}sica, Departamento de Ciencias, Pontificia Universidad Cat\'{o}lica del Per\'{u}, Lima, Peru\\
$^{103}$ Stefan Meyer Institut f\"{u}r Subatomare Physik (SMI), Vienna, Austria\\
$^{104}$ SUBATECH, IMT Atlantique, Nantes Universit\'{e}, CNRS-IN2P3, Nantes, France\\
$^{105}$ Sungkyunkwan University, Suwon City, Republic of Korea\\
$^{106}$ Suranaree University of Technology, Nakhon Ratchasima, Thailand\\
$^{107}$ Technical University of Ko\v{s}ice, Ko\v{s}ice, Slovak Republic\\
$^{108}$ The Henryk Niewodniczanski Institute of Nuclear Physics, Polish Academy of Sciences, Cracow, Poland\\
$^{109}$ The University of Texas at Austin, Austin, Texas, United States\\
$^{110}$ Universidad Aut\'{o}noma de Sinaloa, Culiac\'{a}n, Mexico\\
$^{111}$ Universidade de S\~{a}o Paulo (USP), S\~{a}o Paulo, Brazil\\
$^{112}$ Universidade Estadual de Campinas (UNICAMP), Campinas, Brazil\\
$^{113}$ Universidade Federal do ABC, Santo Andre, Brazil\\
$^{114}$ Universitatea Nationala de Stiinta si Tehnologie Politehnica Bucuresti, Bucharest, Romania\\
$^{115}$ University of Cape Town, Cape Town, South Africa\\
$^{116}$ University of Derby, Derby, United Kingdom\\
$^{117}$ University of Houston, Houston, Texas, United States\\
$^{118}$ University of Jyv\"{a}skyl\"{a}, Jyv\"{a}skyl\"{a}, Finland\\
$^{119}$ University of Kansas, Lawrence, Kansas, United States\\
$^{120}$ University of Liverpool, Liverpool, United Kingdom\\
$^{121}$ University of Science and Technology of China, Hefei, China\\
$^{122}$ University of South-Eastern Norway, Kongsberg, Norway\\
$^{123}$ University of Tennessee, Knoxville, Tennessee, United States\\
$^{124}$ University of the Witwatersrand, Johannesburg, South Africa\\
$^{125}$ University of Tokyo, Tokyo, Japan\\
$^{126}$ University of Tsukuba, Tsukuba, Japan\\
$^{127}$ Universit\"{a}t M\"{u}nster, Institut f\"{u}r Kernphysik, M\"{u}nster, Germany\\
$^{128}$ Universit\'{e} Clermont Auvergne, CNRS/IN2P3, LPC, Clermont-Ferrand, France\\
$^{129}$ Universit\'{e} de Lyon, CNRS/IN2P3, Institut de Physique des 2 Infinis de Lyon, Lyon, France\\
$^{130}$ Universit\'{e} de Strasbourg, CNRS, IPHC UMR 7178, F-67000 Strasbourg, France, Strasbourg, France\\
$^{131}$ Universit\'{e} Paris-Saclay, Centre d'Etudes de Saclay (CEA), IRFU, D\'{e}partment de Physique Nucl\'{e}aire (DPhN), Saclay, France\\
$^{132}$ Universit\'{e}  Paris-Saclay, CNRS/IN2P3, IJCLab, Orsay, France\\
$^{133}$ Universit\`{a} degli Studi di Foggia, Foggia, Italy\\
$^{134}$ Universit\`{a} del Piemonte Orientale, Vercelli, Italy\\
$^{135}$ Universit\`{a} di Brescia, Brescia, Italy\\
$^{136}$ Variable Energy Cyclotron Centre, Homi Bhabha National Institute, Kolkata, India\\
$^{137}$ Warsaw University of Technology, Warsaw, Poland\\
$^{138}$ Wayne State University, Detroit, Michigan, United States\\
$^{139}$ Yale University, New Haven, Connecticut, United States\\
$^{140}$ Yonsei University, Seoul, Republic of Korea\\
$^{141}$  Zentrum  f\"{u}r Technologie und Transfer (ZTT), Worms, Germany\\
$^{142}$ Affiliated with an institute covered by a cooperation agreement with CERN\\
$^{143}$ Affiliated with an international laboratory covered by a cooperation agreement with CERN.\\

\end{flushleft} 
  
\end{document}